\documentclass[lettersize,journal]{IEEEtran}
\usepackage{amsmath,amsfonts}
\usepackage{algorithmic}
\usepackage{array}
\usepackage{textcomp}
\usepackage{stfloats}
\usepackage{url}
\usepackage{verbatim}
\usepackage{graphicx}
\usepackage{array}
\newcolumntype{P}[1]{>{\raggedright\arraybackslash}p{#1}}
\usepackage{stmaryrd}
\usepackage{amsthm}
\usepackage{amssymb}
\usepackage{caption}
\usepackage{courier}
\usepackage{pifont}
\usepackage{color}
\usepackage{verbatim}
\usepackage[hidelinks]{hyperref}
\usepackage[scr=boondoxo]{mathalfa}
\usepackage[section]{placeins}
\usepackage[square,numbers]{natbib}
\graphicspath{ {figures/} }
\usepackage{float}
\usepackage{balance}
\usepackage{breqn}
\allowdisplaybreaks[3]
\def\BibTeX{{\rm B\kern-.05em{\sc i\kern-.025em b}\kern-.08em
    T\kern-.1667em\lower.7ex\hbox{E}\kern-.125emX}}

\newcommand{\G}{\mathcal{G}}
\newcommand{\A}{\mathcal{A}}
\newcommand{\N}{\mathcal{N}}
\newcommand{\V}{\mathcal{V}}
\newcommand{\T}{\mathcal{T}}
\newcommand{\K}{\mathcal{K}}

\usepackage{accents}
\newcommand{\ubar}[1]{\underaccent{\bar}{#1}}
\DeclareMathOperator*{\argmin}{\mathrm{argmin}}
\newcommand{\Exp}[1]{\mathbb{E}\left[#1\right]}

\begin{document}
	\title{Balancing Passenger Transport and Power Distribution: A Distributed Dispatch Policy for Shared Autonomous Electric Vehicles}

\IEEEoverridecommandlockouts
\IEEEpubidadjcol

\author{Jake Robbennolt, Meiyi Li, Javad Mohammadi, Stephen D. Boyles
\thanks{University of Texas at Austin,
Austin, Texas, USA.\\ Contact: {\href{mailto:jr73453@utexas.edu}{\tt jr73453@utexas.edu}}}}

\maketitle

\begin{abstract}
Shared autonomous electric vehicles can provide on-demand transportation for passengers while also interacting extensively with the electric distribution system. This interaction is especially beneficial after a disaster when the large battery capacity of the fleet can be used to restore critical electric loads. We develop a dispatch policy that balances the need to continue serving passengers (especially critical workers) and the ability to transfer energy across the network. The model predictive control policy tracks both passenger and energy flows and provides maximum passenger throughput if any policy can. The resulting mixed integer linear programming problem is difficult to solve for large-scale problems, so a distributed solution approach is developed to improve scalability, privacy, and resilience. We demonstrate that the proposed heuristic, based on the alternating direction method of multipliers, is effective in achieving near-optimal solutions quickly. The dispatch policy is examined in simulation to demonstrate the ability of vehicles to balance these competing objectives with benefits to both systems. Finally, we compare several dispatch behaviors, demonstrating the importance of including operational constraints and objectives from both the transportation and electric systems in the model.  
\end{abstract}

\begin{IEEEkeywords}
Shared Autonomous Electric Vehicles, Grid Resilience, Service Restoration, Alternating Direction Method of Multipliers, Maximum Throughput Dispatch
\end{IEEEkeywords}

\section{Introduction}
Advancements in shared mobility-on-demand services, vehicle automation, and electrification are reshaping the transportation landscape, enabling shared autonomous electric vehicles (SAEVs) to facilitate dynamic interaction between transportation networks and the electric grid \citep{unterluggauer_electric_2022, fagnant_preparing_2015}. Such vehicles have the potential to reduce electricity demand and voltage fluctuation and improve reliability and resilience in the electric grid if charging and discharging schemes are collaboratively optimized \citep{yao_transportable_2019, li_real-time_2021, mohammadi_towards_2023}. However, though many studies on vehicle routing have neglected the impact of electric vehicles (EVs) on the grid by assuming infinite power availability \citep{goeke_routing_2015, pourazarm_optimal_2016}, research has shown that incorrectly managed EV charging could lead to power quality problems \citep{putrus_impact_2009}. Appropriate control can negate these concerns and lead to benefits such as peak shaving and voltage and frequency control, particularly when vehicle-to-grid charging technology is employed \citep{tomic_using_2007}. Furthermore, integrating EVs with renewable energy sources has shown particular promise in helping align electricity demand and supply curves, increasing economic benefits, and improving resilience \citep{lund_integration_2008}. As the SAEV fleet size must be large to serve peak hour demand, the incorporation of constraints related to the energy grid can enable additional uses for these vehicles when they are not needed for transportation. The integration of SAEVs with power grid operations is motivated by both economic and regulatory factors. Grid operators can compensate fleet operators for power support services, creating new revenue streams while avoiding investments in dedicated TESS fleets. During emergencies, regulatory frameworks may require critical infrastructure providers, including SAEV fleets, to support grid restoration efforts. This cooperation benefits both systems -- grid stability ensures SAEV charging availability, while mobile power support helps restore critical loads.

SAEVs can significantly enhance grid resilience through multiple mechanisms. Studies have shown that strategically positioned EVs can provide backup power during outages \citep{reddy_integrating_2024}, support voltage regulation and frequency control through vehicle to grid technology \citep{tomic_using_2007}, enable mobile energy transfer to isolated sections of the grid \citep{yao_transportable_2019}, and facilitate renewable energy integration \citep{mohammadi_strategic_2024, yang_optimal_2024}. The large battery capacity of SAEV fleets makes them particularly valuable for integration with the electric grid \citep{wu_enhancing_2022,yang_optimal_2024}. However, beyond day-to-day operations, there are additional resilience benefits when strains are put on the electric grid \citep{amirioun_resilience-oriented_2023}. Natural disasters can endanger the distribution system, causing equipment failures, blackouts, and even larger scale propagation of failures throughout the network \citep{neumayer_assessing_2013}. While existing research has examined using dedicated transportable energy storage systems (TESSs) for grid restoration \citep{yao_transportable_2019, sun_optimal_2019}, these approaches require costly dedicated fleets that serve only power needs, ignore the continued need for passenger transportation during disasters, do not address privacy concerns between grid operators and vehicle dispatchers, and lack real-time coordination between power and transportation networks. Furthermore, while studies have explored using transit vehicles for grid support \citep{li_resilient_2021}, they focus on fixed-route services and fail to address the unique challenges of coordinating large-scale SAEV fleets that must dynamically balance real-time passenger demands with power distribution needs. These gaps highlight the need for a comprehensive framework that can balance competing demands between passenger service and grid support, preserve privacy between stakeholders, enable real-time implementation, and provide provable passenger throughput guarantees. 

Shared autonomous vehicle (SAV) dispatch is fundamentally a dial-a-ride problem which is typically assumed to encompass a much larger fleet size and have stochastic demand. Many agent based studies have developed routing heuristics to demonstrate the potential value of these fleets \citep{fagnant_travel_2014, levin_general_2017, vosooghi_shared_2019}. However, the range of results suggests a need for an optimization framework that can better control vehicles across scenarios \citep{zhang_control_2016, horl_fleet_2019, hyland_dynamic_2018}. Analytical frameworks used for optimization of the entire fleet can quickly become computationally expensive unless major assumptions are made, particularly when including vehicle rebalancing \citep{robbennolt_maximum_2023} and EV charging \citep{iacobucci_optimization_2019, boewing_vehicle_2020, robbennolt_shared_2024}. Though computationally expensive, these analytical frameworks have been shown to outperform other heuristic models both for standard SAV fleets and SAEV fleets which must charge from the grid \citep{zhang_model_2016,iacobucci_optimization_2019, boewing_vehicle_2020}. Some approaches have developed analytical proofs of maximum throughput which provide additional important service guarantees \citep{kang_maximum-stability_2021, li_real-time_2021, xu_zone-based_2021, robbennolt_maximum_2023}. Finally, innovative approaches for fleet control based on reinforcement learning have been shown to perform well in simulation \citep{skordilis_modular_2022, qin_reinforcement_2022}. However, these methods can lead to less intuitive results and they do not lend themselves well to an analytical characterization of their properties. In contrast, this work builds heavily on analytical approaches with maximum throughput guarantees, extending these frameworks to the context of a cooperative framework integrated with the power distribution system. This ensures the same stability properties while also enabling additional services and enhanced resilience for disaster recovery. 

The contributions of this study are as follows: First, this study introduces a dispatch policy that coordinates the SAEV system with the power distribution network. In our framework, the grid operator manages power dispatch while the vehicle dispatcher optimizes vehicle logistics. Unlike previous research, we acknowledge that even during disruptions to the power grid, vehicles will continue to serve passengers. With this in mind, our proposed model ensures that the integrated efforts of the vehicle dispatcher and power grid operator achieve a balanced approach to maintaining essential transportation services, particularly for critical workers, and supporting end-user power needs during system disruptions \citep{robbennolt_resilience_2023, robbennolt_shared_2024}. Second, this novel cooperative approach allows for joint predictive control and modeling of vehicle and power flows, allowing vehicles to provide services to the electric grid while still maximizing passenger throughput \citep{kang_maximum-stability_2021, li_real-time_2021, xu_zone-based_2021, robbennolt_maximum_2023}. This important property ensures that our policy can serve all passenger demand if any policy can, while also providing ancillary services to the grid. Third, we advance our methodology by developing a novel distributed solution approach that enables local decision-making with minimal communication. The hierarchical approach addresses scalability, stability, privacy, and resilience concerns. This heuristic, based on the alternating direction method of multipliers (ADMM) \citep{boyd_distributed_2011} achieves near-optimal, dynamic solutions quickly. Fourth, our findings from simulations on multiple networks illustrate the ability of the proposed dispatch policy to balance the competing demands of transportation and power grids, ultimately highlighting the critical need to integrate operational constraints and objectives from both sectors into the planning model. The case studies also highlight the computational efficiency of the proposed decomposition approach which is crucial for operationalization. 

\section{Dispatch Policy}
\label{disp}
\begin{figure}[H]
	\centering
	\includegraphics[width=\linewidth]{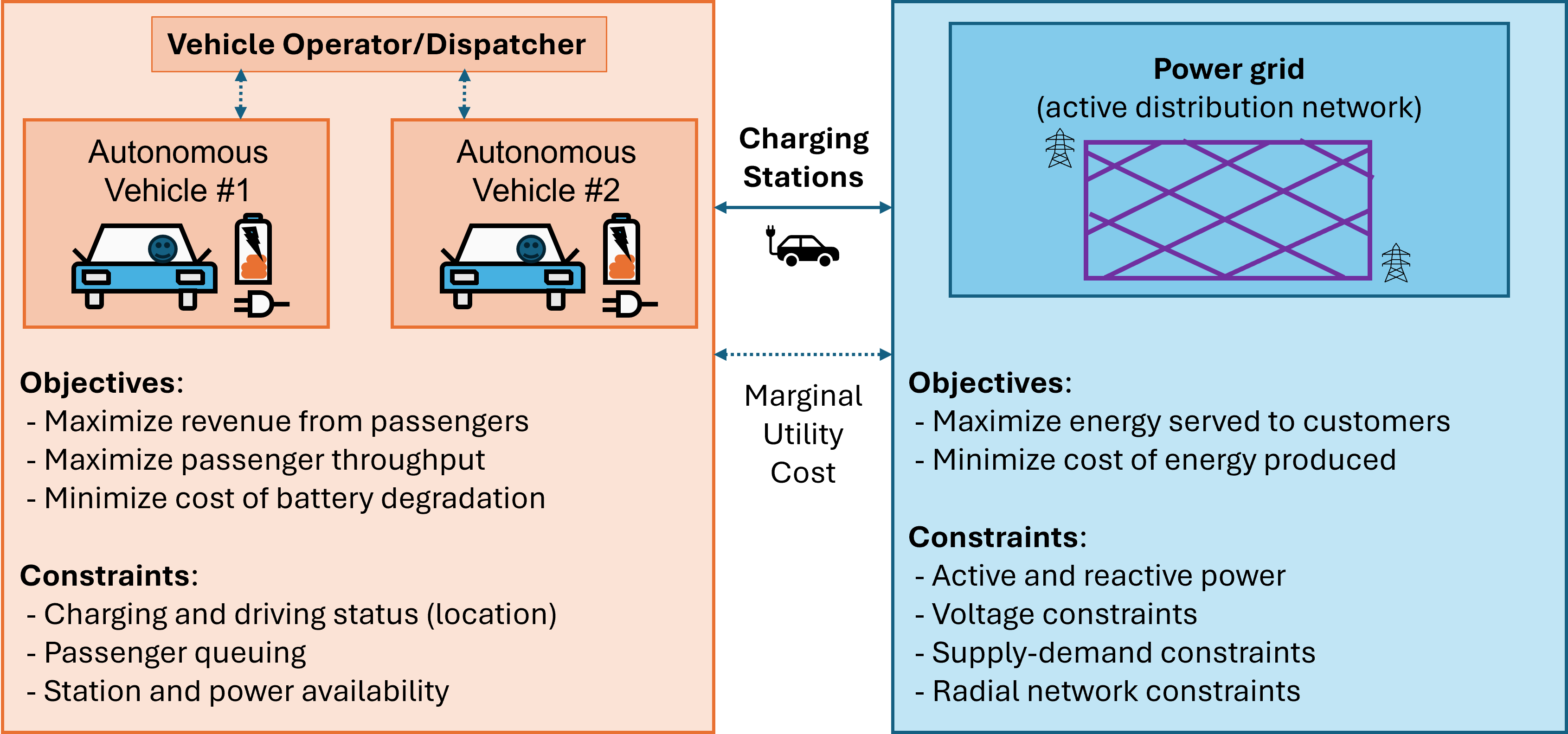}
	\caption{\small Joint optimization between the fleet dispatcher and the electric grid operator.}
	\label{jointinteraction}
\end{figure}

We first consider the power and vehicle dispatch problem as a joint centralized optimization where the vehicle dispatcher and electric grid operator want to maximize their combined profits (see Figure \ref{jointinteraction}). We formulate the problem as a single mixed integer linear program using a model predictive control framework. Within this framework, we must track vehicle movements, passenger and energy flows, and vehicle state of charge over time. This section presents the centralized optimization problem by splitting the constraints into four sets: constraints on passenger queuing, vehicle charging, grid topology, and power flow, before presenting the combined objective. 
Consider a roadway network $\G_R=(\N_R,\A_R)$ and an electric network $\G_E=(\N_E,\A_E)$ with nodes $\N$ and links $\A$. We will use indices $i$ and $j$ to refer to nodes in the electric network and $q$, $r$, and $s$ to refer to nodes in the transportation system. We also define a fleet of vehicles $\V$. We optimize vehicle dispatch through a model predictive control framework. At each timestep $t$, the optimization will run over the time horizon $[t, t+\T]$ (we will use the notation $t_\tau$ as a shorthand to refer to the time $t+ \tau$). Vehicles are controlled by the defined dispatch policy with decisions variables $y_{qr}^v(t_\tau) \in \{0,1\}$ (whether vehicle $v$ will drive between nodes $q$ and $r$), $Y_{qr}^v(t_\tau) \in \{0,1\}$ (whether vehicle $v$ pick up a passenger at $q$ and take them to $r$), $\hat \gamma_{q, ch}^v(t_\tau)\in \{0,1\}$, and $\hat\gamma_{q, dch}^v(t_\tau) \in \{0,1\}$ (whether vehicle $v$ will stay at node $q$ to charge or discharge).

As vehicles can only be dispatched when they have completed a trip, these quantities are only defined when a vehicle is parked at node $q$. We define $x_q^v(t_\tau) \in \{0,1\}$ to denote whether a vehicle is parked at $q$ at time $(t_\tau)$. Based on this setup, the vehicle conservation constraint for each vehicle $v$ is: 
\small
\begin{align}
	\label{decision}
	\sum\limits_{r\in\N_R} y_{qr}^v(t_\tau) + \hat\gamma_{q, ch}^v(t_\tau) + \hat\gamma_{q, dch}^v(t_\tau) \nonumber \leq x_q^v(t_\tau), \\ \qquad \forall q \in \N_R, \forall v \in \V, \forall t_\tau \in \T.
\end{align}
\normalsize
Each vehicle can only choose a single action (drive, charge, or discharge) each time they are available for dispatch.  

\subsection{Passenger Queuing Constraints}
\label{sectqueue}
Passengers are tracked based on a queuing model, and vehicle and passenger movements are tracked across the time horizon. We can constrain the dispatch of SAEVs traveling from $q$ to $r$ depending on whether vehicle $v$ is actually parked at node $q$. To do this we define the exogenous travel time $C_{qs}$ to be the travel time between nodes $q$ and $s$. This travel time is assumed to be constant (not impacted by SAEV dispatch) but could include congestion external to the dispatch problem. Then, vehicle state evolves as:
\small
\begin{align}
	x_q^v(t_\tau + 1) = x_q^v(t_\tau) + &\sum_{s \in \N_R} y_{sq}^v(t_\tau + 1 - C_{sq}) - \sum_{r \in \N_R} y_{qr}^v(t_\tau), \nonumber \\ &\qquad \forall q \in \N_R, \forall v \in \V, \forall t_\tau \in \T. \label{Constr_x}
\end{align}
\normalsize
In general vehicles travel to the passenger pickup location and then take them to their destination, so the number of vehicles carrying passengers can be less than the total number of vehicles driving between nodes: 
\small
\begin{align}
	Y_{qr}^v(t_\tau) &\leq y_{qr}^v(t_\tau), && \forall (q,r) \in \N_R^2, \forall v \in \V, \forall t_\tau \in \T. \label{eqy}
\end{align}
\normalsize
We consider exogenous demand $d_{qr}(t)$ to enter the network at node $r$ with destination $s$ at time $(t)$. These passengers form a separate queue $w_{qr}(t_\tau)$ at each origin for each destination. Since the demand is unknown for all $\tau > 0$, queues evolve within the model predictive control framework based on \textit{predicted} future demand ($\tilde{d}_{qr}(t_\tau)$). Then, $w_{qr}(t_\tau)$ evolves as follows: 
\small
\begin{align}
	w_{qr}(t+1) &= w_{qr}(t_\tau) + \tilde{d}_{qr}(t_\tau) - \sum_{v \in \V}Y_{qr}^v(t_\tau), \nonumber  \\& \forall (q,r) \in \N_R^2, \forall t_\tau \in \T. \label{eqw}
\end{align}
\normalsize
Finally, the number of vehicles dispatched to serve passengers must be less than to total passenger demand:
\small
\begin{align}
	\sum_{v \in \V}Y_{qr}^v(t_\tau) \leq w_{qr}(t_\tau), && \forall (q,r)\in\N_R^2, \forall t_\tau \in \T.
	\label{Equation_2}
\end{align}
\normalsize

\subsection{Vehicle Charging Constraints}
$\hat\gamma_{q, ch}^v(t_\tau) \in \{0,1\}$ and $\hat\gamma_{q, dch}^v(t_\tau) \in \{0,1\}$ define whether vehicles are charging or discharging, but not how much power is being transferred. To track power flows, we define $\gamma_{q, ch}^v(t_\tau) \in[0,1]$ and $\gamma_{q, dch}^v(t_\tau) \in[0,1]$ and maximum charging and discharging rates $\Gamma^v_{q,ch}$ and $\Gamma^v_{q,dch}$. We constrain power flows based on charging status and calculate the actual power taken from the grid based on the power flows:
\small
\begin{align}
	\label{chargedecision}
	\gamma_{q, ch}^v(t_\tau) \leq \hat\gamma_{q, ch}^v(t_\tau),  && \forall q \in \N_R, \forall v \in \V, \forall t_\tau \in \T,\\
	\label{dischargedecision}
	\gamma_{q, dch}^v(t_\tau) \leq  \hat\gamma_{q, dch}^v(t_\tau),  && \forall q \in \N_R, \forall v \in \V, \forall t_\tau \in \T, 
\end{align}
\begin{align}
	\label{ep_def}
	ep^v_q(t_\tau) &= \gamma_{q, ch}^v(t_\tau)\Gamma^v_{q,ch} - \gamma_{q, dch}^v(t_\tau)\Gamma^c_{q,dch},\nonumber  \\& \forall q\in\N_R, \forall v\in \V \forall t_\tau \in \T, 
\end{align}
\normalsize
where $ep^v_q(t_\tau)$ and $eq^v_q(t_\tau)$ denote the active and reactive power that any individual vehicle parked at node $q \in N_R$ takes from or gives to the grid. As in \citet{singh_electric_2020}, we assume that EVs are equipped with bi-directional chargers that can inject/absorb reactive power without affecting the state of charge or battery life. Based on the active power taken from the grid, each vehicle is limited in the amount of reactive power that they can take \citep{kisacikoglu_single-phase_2015}. We can then constrain the reactive power flow based on a linear approximation of the real quadratic constraint (where $\bar{es}_{q}^v$ is the maximum allowed real power flow):
\small
\begin{align}
	\label{eSquare1}
    ep^v_{q}(t_\tau)^2 &+ eq_{q}^v(t_\tau)^2 \leq ([\hat \gamma_{q,ch}^v(t_\tau) + \hat \gamma_{q,dch}^v(t_\tau) ]\bar{es}_{q}^v)^2, \nonumber  \\& \forall q\in\N_R, \forall v\in \V \forall t_\tau \in \T.
\end{align}
\normalsize

Recall that if $[\hat \gamma_{q,ch}^v(t_\tau) + \hat \gamma_{q,dch}^v(t_\tau)] = 0$, then no active or reactive power is allowed to be transferred through the charging station at $q$. The active power is already constrained by constraint \eqref{ep_def}, but this term constrains the reactive power as well. The linearization process is commonly used when constraining power flow though the distribution system \citep{baran_optimal_1989,turitsyn_distributed_2010,yeh_adaptive_2012,yao_transportable_2019,li_resilient_2021}. 

Next, we need to ensure that any charging and discharging by vehicles at nodes in the roadway network is transferred to the electric grid. We define $EP_i(t_\tau)$ and $EQ_i(t_\tau)$ to be the amount of active and reactive power vehicles take (positive) or give (negative) to the grid at node $i$. The power distribution and transportation networks are connected by charging stations located at nodes, with connections denoted using the binary variable $\delta_{qi} \in \{0,1\}$ where $i$ is in the electric grid and $q$ is in the roadway network. Then, the values of power for individual vehicles need to be aggregated and converted to demands or supplies on the electric grid:
\small
\begin{align}
	\label{EPR}
	EP_i(t_\tau) &= \sum_{q\in N_R} \sum_{v \in \V} \delta_{qi} ep^v_{q}(t_\tau), &&  \forall i\in\N_E, \forall t_\tau \in \T,\\
	\label{EQR}
	EQ_i(t_\tau) &= \sum_{q\in N_R} \sum_{v \in \V} \delta_{qi}eq^v_{q}(t_\tau), &&  \forall i\in\N_E, \forall t_\tau \in \T.
\end{align}
\normalsize
We also need to ensure that vehicle are only allowed to charge and discharge if stations are available that their current node: 
\small
\begin{align}
	\label{NQ}
	\sum_{v \in \V} \left[\hat\gamma_{q,ch}^v(t_\tau) \right. + \left. \hat \gamma_{q,dch}^v(t_\tau) \right] \leq N_q,  && \forall q\in\N_R, \forall t_\tau \in \T,
\end{align}
\normalsize
where $N_q$ is the number of charging stations at $q$. 

For the vehicle energy tracking, we calculate the energy impacts of the dispatch decision as soon as vehicles are dispatched. Denote $e^v(t_\tau)$ as the charge of vehicle $v$ at time $(t_\tau)$, which can be updated as: 
\small
\begin{align}
	&e^v(t_\tau + 1) = e^v(t_\tau) - \sum\limits_{(q,r)\in\N_R^2} y_{qr}^v(t_\tau)B_{qr} \nonumber \\ & + \sum\limits_{q \in \N_R} \left[ \gamma_{q,ch}^v(t_\tau)\Gamma^v_{q,ch}\eta^v_{q,ch} - \frac{\gamma_{q,dch}^v(t_\tau)\Gamma^v_{q,dch}}{\eta^v_{q,dch}} \right],  \nonumber \\ & \qquad \qquad \qquad \qquad \forall v \in \V, \forall t_\tau \in \T, \label{gamma}
\end{align}
\normalsize
where $B_{qs}$ is the energy requirements for traveling between $q$ and $s$. 

Once each vehicle's state of charge is known, it is straightforward to constrain it between some upper ($\bar e^v$) and lower ($\ubar e^v$) bound:
\small
\begin{align}
	\label{ebounds}
	\ubar e ^v \leq e^v(t_\tau) \leq \bar e ^v,  && \forall v \in \V, \forall t_\tau \in \T.
\end{align}
\normalsize
We assume $\ubar e^v$ is calculated based on the energy consumption needed to get to the nearest charging station. 

\subsection{Grid Topology Constraints}
The LinDistFlow power flow model described in Section \ref{powerFlow} assumes a radial network, so the distribution system may need to be reconfigured \citep{li_resilient_2021, xin_rolling_2022}. This problem is coupled with the vehicle dispatch problem since changing power demands at charging stations can affect the pattern of flows in the distribution system, necessitating a different network topology. This coupling is particularly important during service restoration when SAEVs must be strategically positioned to support isolated sections of the grid. The variable $u_{ij}(t_\tau) \in \{0,1\}$ denotes the line state; $u_{ij}(t_\tau) = 0$ if a line has failed. We also define $s_{ij}(t_\tau) \in \{0,1\}$ to indicate whether power flow should be allowed to flow. $s_{ij}^d$ and $s_{ji}^r$ are defined similarly for directed line flow between $(i,j)$ and $(j,i)$ respectively. Since power may only flow in one direction on each line, the topology is constrained by:
\small
\begin{align}
	\label{damage}
	s_{ij}(t_\tau) \leq u_{ij}(t_\tau), &&\forall (i,j) \in \A_E, \forall t_\tau \in \T,\\
	\label{sum}
	s_{ij}(t_\tau)  = s_{ij}^d(t_\tau) + s_{ji}^r(t_\tau),  && \forall (i,j) \in \A_E, \forall t_\tau \in \T,\\
	\label{origin0}
	s_{j0}^r(t_\tau) = 0, &&\forall j \in \K_G \cup \K_C, \forall t_\tau \in \T,\\
	\label{origin1}
	s_{0j}^d(t_\tau) = 1, &&\forall j \in \K_G, \forall t_\tau \in \T,\\
	\label{origin2}
	s_{0j}^d(t_\tau) \leq 1, &&\forall j \in \K_C, \forall t_\tau \in \T,
\end{align}
\normalsize
where the set $\K_G$ is the set of source nodes connected to the grid. $\K_C$ is the set of EV charging stations that can act as source nodes if disconnected from the grid but need not otherwise (the subset of $\N_E$ that is attached to a charging station in $\N_R$). Here, constraint \eqref{damage} enforces the limitation that power cannot flow on damaged lines. Constraint \eqref{sum} ensures that power can only flow in one direction. Finally, constraints \eqref{origin0} and \eqref{origin1} require power flows from the virtual supersource while constraints \eqref{origin0} and \eqref{origin2} allow power flows from the virtual supersource. 

We can also define the variable $z_i(t_\tau) \in \{0,1\}$ to represent the state of power supply of the distribution node. These variables, along with virtual demands $f_i^L = 1$ for all nodes and virtual power flows $f_{ij}$ allows us to construct the radial network:
\small
\begin{align}
	\label{sum1}
	\sum_{i \in \N_E : (i,k) \in \A_E} s_{ik}^d(t_\tau) &+ \sum_{j \in \N_E : (j,k) \in \A_E} s_{jk}^r(t_\tau) \leq 1,  \nonumber \\ & \forall k \in N_E, \forall t_\tau \in \T,\\
	\label{sumz}
	\sum_{i \in \N_E : (i,k) \in \A_E} s_{ik}^d(t_\tau) &+ \sum_{j \in \N_E : (j,k) \in \A_E} s_{jk}^r(t_\tau)   \geq z_k(t_\tau),  \nonumber \\ & \forall k \in N_E ,\forall t_\tau \in \T,\\
	\label{nodeline}
	s_{ij}(t_\tau) - 1 &\leq z_i(t_\tau) - z_j(t_\tau) \leq 1-s_{ij}(t_\tau),  \nonumber \\ & \forall (i,j) \in \N_E^2, \forall t_\tau \in \T,\\
	\label{ZFL}
	z_i(t_\tau) &= \sum_{j \in \N_E} f_{ji}(t_\tau) - \sum_{j \in \N_E} f_{ij}(t_\tau),  \nonumber \\ &  \forall i \in \N_E, \forall t_\tau \in \T,\\
	\label{N}
	-s_{ij}(t_\tau)N &\leq f_{ij}(t_\tau) \leq s_{ij}(t_\tau)N,  \nonumber \\ & \forall (i,j) \in \A_E, \forall t_\tau \in \T.
\end{align}
\normalsize
Constraint \eqref{sum1} requires the in-degree of each node to be less than or equal to 1 to ensure the network is radial. Constraint \eqref{sumz} ensures that each child node (powered) is assigned a single parent node. Constraint \eqref{nodeline} enforces consistency between the state of each line and the state of the node on either side (if node $i$ is powered by a source and $j$ is not then the line between them cannot allow power to flow). Constraints \eqref{ZFL} and \eqref{N} constitute flow conservation on the virtual network where $N = |\N_E|$ is the number of distribution nodes. 

\subsection{Power Flow Constraints}
\label{powerFlow}
To model the power flow we use the LinDistFlow model \citep{turitsyn_distributed_2010, yeh_adaptive_2012, li_resilient_2021} and assume that any fractional amount of load can be served. This is done using the variable $l_i(t_\tau) \in [0,1]$. Based on network topology we can define:
\small
\begin{align}
	\label{LZConst}
	l_i(t_\tau) \leq z_i(t_\tau), &&\forall i \in \N_E, \forall t_\tau \in \T.
\end{align}
\normalsize
We can then define active and reactive power constraints to ensure we serve all demand ($P^L_i(t_\tau)$ and $Q^L_i(t_\tau)$ respectively) that has been chosen for pickup. 
\small
\begin{align}
	\label{Pbalance}
	\sum_{j \in \N_E : (i,j) \in \A_E} &P_{ij}(t_\tau) = P^G_i(t_\tau) - l_i(t_\tau)P^L_i(t_\tau)   - EP_i(t_\tau), \nonumber \\ &\qquad \forall i \in N_E, \forall t_\tau \in \T,\\
	\label{Qbalance}
	\sum_{j \in \N_E : (i,j) \in \A_E} &Q_{ij}(t_\tau) = Q^G_i(t_\tau)  - l_i(t_\tau)Q^L_i(t_\tau)   - EQ_i(t_\tau), \nonumber \\ & \qquad \forall i \in N_E, \forall t_\tau \in \T.
\end{align}
\normalsize
The generated power $P^G_i(t_\tau)$ and $Q^G_i(t_\tau)$ is constrained within known limits:
\small
\begin{align}
	\label{Pbound}
	\ubar P_{i}^G \leq P_{i}^G(t_\tau) \leq \bar P_{i}^G,  && \forall (i,j) \in \A_E, \forall t_\tau \in \T,\\
	\label{Qbound}
	\ubar Q_{i}^G \leq Q_{i}^G(t_\tau) \leq \bar Q_{i}^G,  && \forall i \in \N_E, \forall t_\tau \in \T.
\end{align}
\normalsize
Next, the active and reactive power flows ($P_{ij}(t_\tau)$ and $Q_{ij}(t_\tau)$) must be bounded based on the real power constraint. This is done using the same linear approximation as above for the constraint (where $\bar S_{ij}$ is the maximum allowed real power): 
\small
\begin{align}
	\label{Square1}
    P_{ij}(t_\tau)^2 + Q_{ij}(t_\tau)^2 \leq s_{ij}(t_\tau) \bar S_{ij}^2,  && \forall (i,j) \in \A_E, \forall t_\tau \in \T.
\end{align}
\normalsize
Constraints \eqref{V1} and \eqref{V2} relate the voltage ($V_i(t_\tau)$) to the power flows using the resistance $R_{ij}$ and reactance $X_{ij}$ of the lines. Then, constraints \eqref{V3} and \eqref{V4} set a constant voltage $V_0$ for all nodes that supply power, and constrain the voltage between set bounds ($\ubar V_i$ and $\bar V_i$) for all other nodes.

\small
\begin{align}
	\label{V1}
	V_i(t_\tau) - V_j(t_\tau) \leq M\left( 1-s_{ij}(t_\tau) \right) & + \frac{R_{ij}P_{ij}(t_\tau) + X_{ij}Q_{ij}(t_\tau)}{V_0},   \nonumber \\ &  \forall (i,j) \in \A_E, \forall t_\tau \in \T,\\
	\label{V2}
	V_i(t_\tau) - V_j(t_\tau) \geq M\left(s_{ij}(t_\tau) -1\right)& + \frac{R_{ij}P_{ij}(t_\tau) + X_{ij}Q_{ij}(t_\tau)}{V_0},   \nonumber \\ &  \forall (i,j) \in \A_E, \forall t_\tau \in \T,\\
	\label{V3}
	V_i(t_\tau) = V_0, \qquad & \forall i \in \K_G, \forall t_\tau \in \T,\\
	\label{V4}
	z_i(t_\tau) \ubar V_i \leq V_i(t_\tau) \leq z_i(t_\tau) \bar V_i,  \qquad &  \forall i \in \N_E\setminus \K_G, \forall t_\tau \in \T.
\end{align}
\normalsize

\subsection{Objective Function}
\label{objectivefunction}
In an ideal scenario, the dispatch policy ($\pi^\star$) would serve all passenger and electricity demand at every timestep. In this case, the goal should be to minimize either the total electricity consumption or the cost of electricity. However, this is unlikely to be the case in a disrupted network where vehicles may have competing demands from passengers and electric loads. Therefore, in this paper we \textit{maximize} the profit or social welfare (which will also minimize electricity consumption/cost where possible). To achieve this, we add up revenues from serving passengers and electric loads and subtract the cost of generating electricity and the battery degradation cost of the batteries. The objective function is as follows: 
\small
\begin{align}
	\label{objective}
	&\frac{1}{\T} \sum_{\tau = 1}^\T \biggl[               \sum_{(r,s)\in \N_R^2} \mu_0(r,s,t) w_{rs}(t) \sum_{v \in \V} Y_{rs}^v(t_\tau)   \nonumber 
	\\&+ \sum_{(r,s)\in \N_R^2} \mu_1(r,s,t_\tau) \sum_{v \in \V}Y_{rs}^v(t_\tau)   \nonumber 
	\\& + \sum_{i \in \N_E} \mu_2(i,t_\tau)l_i(t_\tau) P_i^L(t_\tau) - \sum_{i \in \N_E} \mu_3(i,t_\tau) P_i^G(t_\tau)\nonumber  
	\\&-  \sum_{v\in\V} \mu_4(v) \sum_{(q)\in \N_R} \left( \gamma_{q,ch}^v(t_\tau)\Gamma_{ch} + \gamma_{q,dch}^v(t_\tau)\Gamma_{dch} \right)  
	\biggr].
\end{align}
\normalsize
In the objective \eqref{objective}, the $\mu \geq 0$ functions are pricing functions set by the operator. The term $\mu_0$ is a demand responsive cost that should increase the payments by customers as demand in their queue increases. Term $\mu_1$ represents revenue from passengers once they are dropped at their destination based on distance or average travel times. Term $\mu_2$ represents revenue generated by payments for energy supplied to customers. Term $\mu_3$ is the price of generating energy. Finally, $\mu_4$ is the cost of battery degradation from charging and discharging. Since the initial queue lengths $w_{rs}(t)$ are constant with respect to $\pi^\star$, the final optimization problem is a mixed-integer linear problem (MILP), assuming linear cost functions $\mu$.

An important property of this objective function is that it ensures stability within the transportation network if it is possible (i.e. the expected number of waiting passengers remains bounded over time). The network is stable if there exists a $\kappa<\infty$ such that
$    \lim\limits_{T\rightarrow\infty} \frac{1}{T} \sum\limits_{t=1}^T \sum\limits_{(r,s)\in\N^2} \Exp{w_{rs}(t)} \leq \kappa$.

The objective function proposed in \citet{kang_maximum-stability_2021} for SAV dispatch is proven to stabilize demand if it is possible to do so. Within this electric vehicle framework the same is true. That proof relies on the first objective term (the demand responsive price) increasing to infinity if queues increase to infinity. This will eventually prioritize passenger service more than providing power to the electric grid and will ensure vehicles serve the longest queues first to maintain stability. For a more extensive discussion of this property and the impacts of electrification on stable dispatch, see \citet{robbennolt_resilience_2023}.

\section{Distributed Dispatch}
The model predictive control algorithm developed in the previous section is beneficial because it allows coordination between the power system and transportation system. Though this coordination is advantageous, it can make the problem substantially harder than solving each problem separately. In addition, though previous optimal dispatch strategies developed maximum throughput policies that did not need a time horizon and only tracked aggregate vehicle flows \citep{li_real-time_2021, robbennolt_maximum_2023}, the need to track state of charge and provide real-time grid services means that individual vehicles must be tracked over a time horizon. Each of these complications increases the size of the problem, adding additional complexity. 

\begin{figure}[h]
	\centering
	\includegraphics[width=\linewidth]{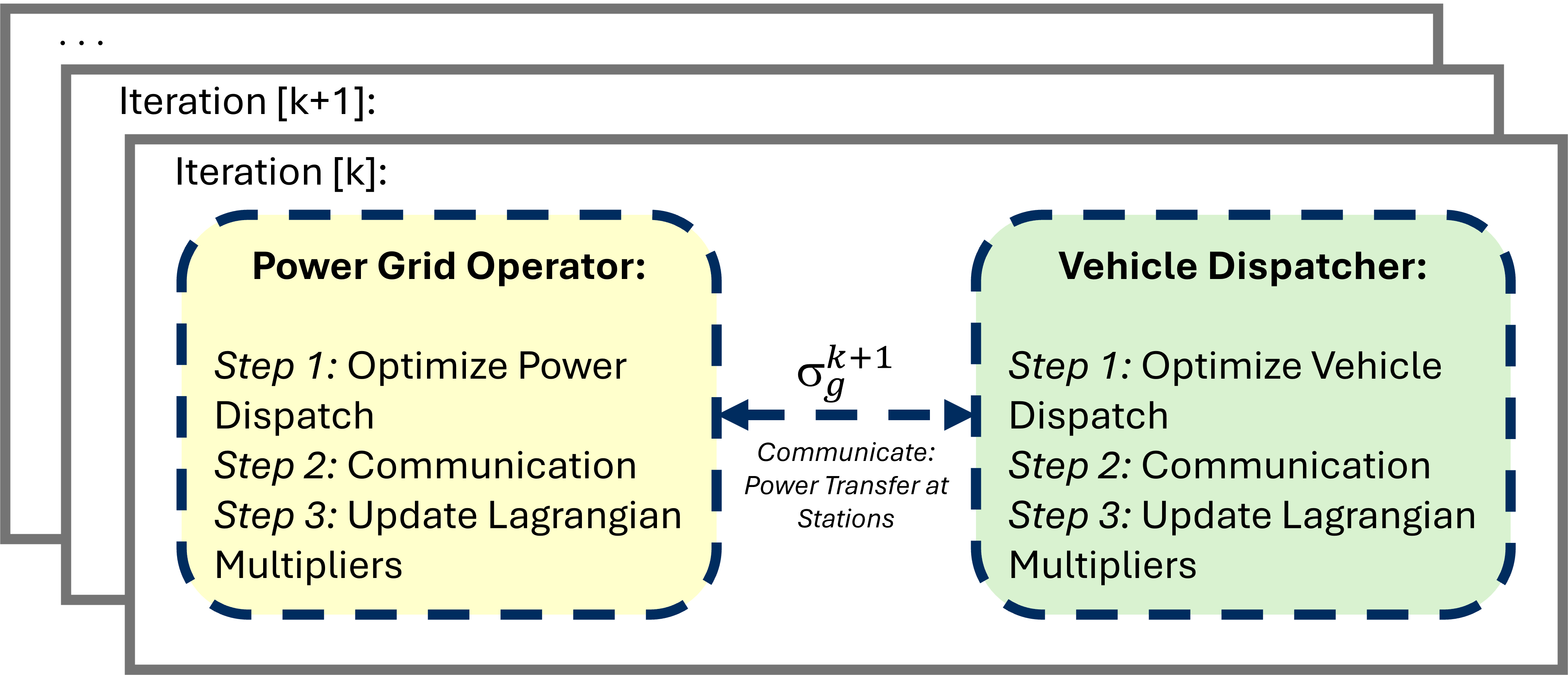}
	\caption{\small Communications between power distribution system operator and SAEV dispatcher.}
	\label{ADMM_upper}
\end{figure}

In this section, we propose a distributed solution method which is scalable even for long time horizons and large fleet sizes and which also preserves privacy. In the decentralized algorithm, the vehicle dispatcher and power grid operator each solve their own problems, passing power flow estimates at charging stations and iterating until consensus (Figure \ref{ADMM_upper}). Section \ref{sect:veh} will demonstrate that to solve the dispatch sub-problem, each vehicle will also solve its own lower-level optimization, passing passenger pickup and charging information to the vehicle dispatcher and iterating until convergence (Figure \ref{ADMM_lower}). Separating the problem into this hierarchical structure allows privacy to be preserved at multiple levels (i.e. the privacy of each vehicle's decisions as well as the privacy of the dispatcher and grid operator). In addition, it enables computational effort to be focused on the vehicle dispatch subproblem which is harder to solve. Finally, this decomposition approach allows for large scale parallelization and could potentially allow the time-scales of the vehicle dispatch and power flow optimization problems to be decoupled in the future. 

The distributed solution method proposed is based on ADMM, a well-established distributed optimization technique. Though ADMM cannot guarantee convergence of MILPs, it is a useful heuristic since it allows the problem to be decomposed. At the upper level, while the integer variables defining dispatch remain fixed and the grid remains in a single radial configuration, the goal is to determine how much power SAEVs should take from the grid. This problem becomes a linear program (convex) when the integer routing and radial variables are fixed, so ADMM will converge exactly. While vehicles are switching routes, the discontinuities can cause issues with convergence. However, as the number of vehicles grows, these effects diminish, and the approximation improves. At the same time, as the problem grows (especially as the time horizon gets longer and the number of vehicles grows) the centralized problem becomes much more difficult to solve and the computational benefits of each vehicle optimizing its own route become more pronounced. 

Although there are other potential solution algorithms for MILPs such as optimal condition decomposition or Benders decomposition, we believe that the proposed heuristic has substantial benefits. In particular, the hierarchical structure of ADMM allows for privacy preservation at multiple levels and could allow future decoupling of the time-scales of the problem \citep{candas_comparative_2020}. In addition, though other methods might offer stronger convergence guarantees, the extremely large number of binary variables could make their application more challenging and they lack the high parallelization potential of the distributed ADMM architecture. We will demonstrate in the case studies below that the unique problem structure allows for very fast near-optimal solutions using this specially tailored ADMM approach. 

Many of the constraints in section \ref{disp} are separable by system (roadway or electric network). However, constraints \eqref{EPR} and \eqref{EQR} aggregate the charge taken by vehicles from the grid at the charging stations and link the vehicle dispatch and power flow problems. In order to create an algorithm to split the decisions of the vehicle dispatcher from the decisions of the power grid operator, each system can create a local copy of the variables $EP_i(t_\tau)$ and $EQ_i(t_\tau)$ which the vehicle dispatcher uses in  \eqref{EPR} and \eqref{EQR} and the power grid operator uses in \eqref{Pbalance} and \eqref{Qbalance}. Based on this idea, we rewrite the centralized problem as:
\small
\begin{subequations}
    \label{ADMM_optim}
    \begin{flalign}
        \label{obj}
		\min_{\sigma} \qquad \text{Separable Objective: $f^R(\sigma) + f^E(\sigma)$}, &&
    \end{flalign}
    \begin{flalign}
        \mathrm{s.t.} \qquad & \text{Separable Constraints R and E}, \\
        & \sigma^R - \mu = 0, \qquad \sigma^E - \mu = 0, && 
    \end{flalign}
\end{subequations}
\normalsize
where $\sigma$ represents the estimate of the aggregate charge that vehicles take from the grid (a vector of all $EP_i(t_\tau)$ and $EQ_i(t_\tau)$). $\mu$ represents the consensus term that these estimates should converge to. The separable constraints are equations \eqref{decision} -- \eqref{ebounds} for the roadway network (R) and equations \eqref{damage} --  \eqref{V4} for the power grid (E). This problem takes the form of a \textit{consensus} ADMM problem with two agents (the power grid operator and the vehicle dispatcher) where $EP_i(t_\tau)$ and $EQ_i(t_\tau)$ have been replaced by their local copies $EP^R_i(t_\tau)$ and $EQ^R_i(t_\tau)$ in equations \eqref{EPR} and \eqref{EQR}. In equations \eqref{Pbalance} and \eqref{Qbalance} they have been replaced by local copies $EP^E_i(t_\tau)$ and $EQ^E_i(t_\tau)$. No other variables are shared between the two agents, so no other equations must be modified. 

We will define the index $g$ to represent either the roadway $R$ or electric networks $E$ to simplify notation. Define $\lambda_E^g$ as the Lagrangian multiplier associated with the consensus constraint and $\rho_E$ as the ADMM penalty parameter. We refer to this penalty as $\rho_E$ because it is associated with convergence of charging patterns based on fluctuations in the electric grid, though this parameter is used by the vehicle dispatcher as well to determine the relative benefit of providing power. The reason for this distinction will become clear in the next section. We also define the scaled dual variable $u^g = \frac{\lambda^g}{\rho_E}$, and the average of the two power flow estimates $\bar{\sigma}$. Then, this formulation simplifies to the iterative procedure of solving for $\sigma^g$ and then updating the scaled dual variables (see \citet{boyd_distributed_2011}): 
\small
\begin{align}
    \sigma_{k+1}^g = &\argmin_{\sigma} \biggl[ f^g(\sigma^g)
    + \frac{\rho_E}{2} \| \sigma^g - \bar\sigma_k + u_k^g \|_2^2 \biggr], \nonumber \\
    &\mathrm{s.t.} \qquad  \text{Separable Constraints $g$}, \label{final_ADMM1}
\end{align}
\begin{align}
    u^g_{k+1} = u^g_k + \sigma^g_{k+1} - \bar{\sigma}_{k+1}.
    \label{final_ADMM2}
\end{align}
\normalsize
The mixed-integer quadratic problems (MIQPs) \eqref{final_ADMM1} can be solved separately by the vehicle dispatcher and power distribution system operator. Each system will determine optimal vehicle movements and power flows (respectively), and both systems will estimate the optimal power flow at charging stations. At each iteration $k$, only the information about predicted power flows at charging stations will be communicated, allowing each system to independently update their own Lagrangian multipliers and continue to the next iteration until convergence. 

\subsection{Vehicle Dispatch Subproblem}
\label{sect:veh}
The upper-level optimization results in two sup-problems, one for power flow optimizations and one for vehicle dispatch. We do not further simplify the power flow problem as it is generally easier to solve. However, the vehicle dispatch sup-problem has many integer variables and remains difficult to solve quickly. 

Based on the ADMM formulation in equation \eqref{final_ADMM1} and \eqref{final_ADMM2}, we now formulate the vehicle dispatch subproblem (which must be solved at each iteration). The decision variables which relate vehicles to each other are the dispatch decisions to pick up passengers ($Y$), as well as the decisions about the amount of power to take from the grid ($ep$ and $eq$) and where to charge or discharge ($\hat{\gamma_{ch}}$, and $\hat{\gamma_{dch}}$), (referred to in aggregate as the vector $z^v$ for each vehicle). Then, the lower-level vehicle dispatch problem can be written as:
\small
\begin{subequations}
    \label{VD_ADMM}
    \begin{flalign}
        \min_{z,\mu} \qquad \sum_{v\in\V} f^v(z^v) + g\left(\sum_{v\in\V}z^v\right),&&
    \end{flalign}
    \begin{flalign}
        \mathrm{s.t.} \qquad & \text{Separable Constraints } v, &&\forall v \in \V, \label{v_all} \\
        & h\left(\sum_{v\in \V} z^v\right) \leq \omega, \label{others}
    \end{flalign}
\end{subequations}
\normalsize
where constraint \eqref{v_all} includes all of the separable constraints that can be split between vehicles (equations \eqref{decision}, \eqref{chargedecision}, \eqref{dischargedecision}, \eqref{Constr_x}, \eqref{eqy}, \eqref{ep_def}, \eqref{eSquare1},  \eqref{gamma}, and \eqref{ebounds}. Constraint \eqref{others} incorporates the non-separable constraints \eqref{eqw}, \eqref{Equation_2}, and \eqref{NQ}. Constraints \eqref{EPR} and \eqref{EQR} can be dropped since we calculate the aggregate charge directly in the objective function. The first term in the objective is the separable constraints for the roadway network, each of which can also be separated for each vehicle. The second term is the function added by the upper-level ADMM, which requires consensus with the electric grid, so must account for the sum of all charging behaviors.

\begin{figure}[h]
	\centering
	\includegraphics[width=\linewidth]{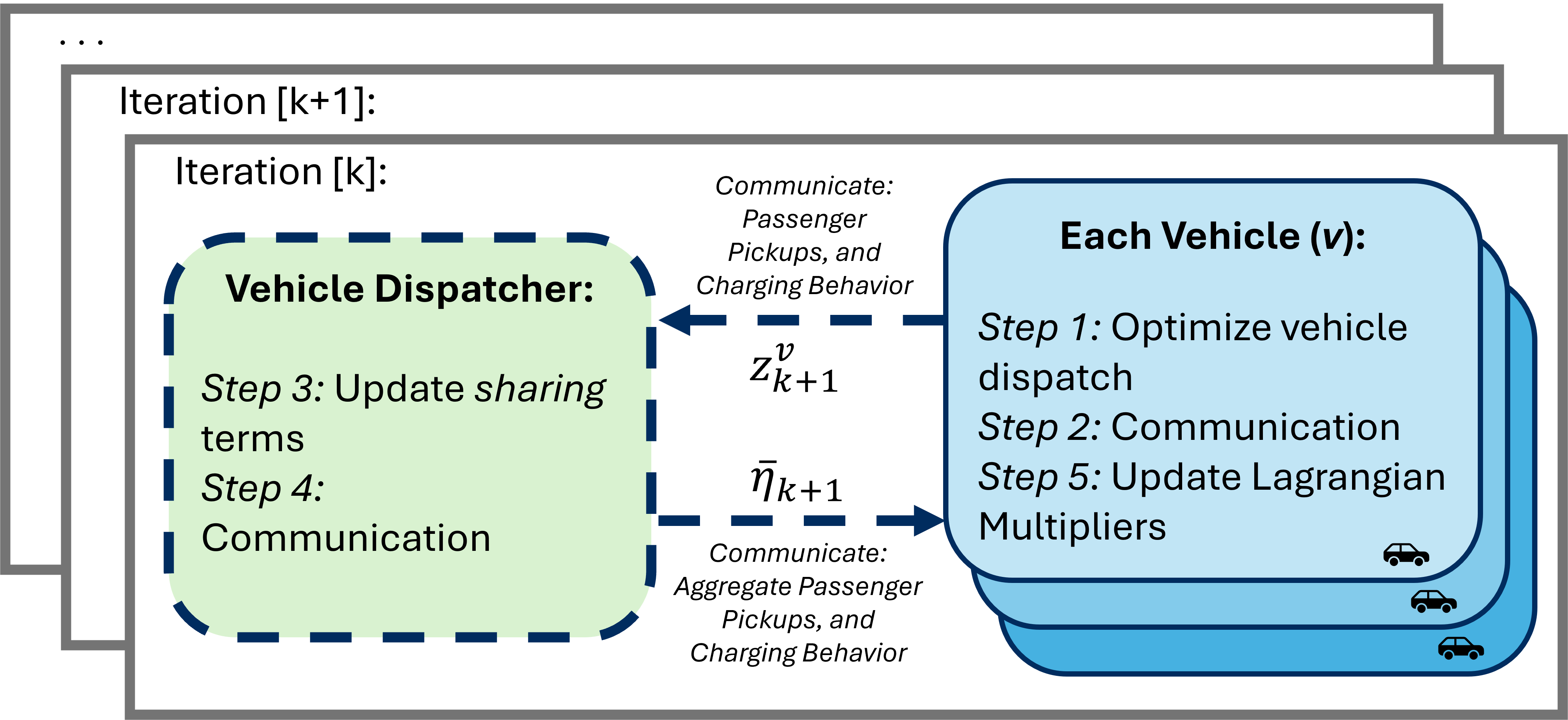}
	\caption{\small Communications between vehicles and central controller.}
	\label{ADMM_lower}
\end{figure}

Note that this MILP includes a large number of integer variables, and the size of the problem still increases rapidly as the size of the roadway network increases. Further, this problem must be solved many times before the solution converges to the solution of problem \eqref{objective}, and the dispatch problem \eqref{objective} must be solved at every timestep (every 15 seconds -- 1 minute). This means the vehicle dispatch subproblem must be solved very quickly. To achieve a sufficiently fast solution we will again propose an ADMM approach, this time distributing subproblems to each vehicle while retaining a the vehicle dispatcher as a centralized coordinator (see figure \ref{ADMM_lower}). The inclusion of a term of the form $\sum_{v \in \V} z^v_{qr}$ in the objective and constraints makes this problem a \textit{sharing} problem which can be written in ADMM form by making a local copy of all shared variables \citep{boyd_distributed_2011}. To do this, replace $z^v$ in optimization problem \eqref{VD_ADMM} with local copies $\eta^v$ and then add the consensus constraint $ z^v - \eta^v = 0$, $\forall v \in \V$.
Define $\rho_R$ as the ADMM penalty parameter, $\lambda^v$ as the Lagrange multipliers associated with ADMM constraint. As above, $u^v$ is the scaled form of the Lagrangian multiplier $\lambda^v$ ($u^v = \frac{\lambda^v}{\rho_R})$. Also, based on the formulation by \citet{boyd_distributed_2011}, we replace $\eta^v$ with $\bar \eta$ for efficiency, Then, we can write the scaled form of ADMM as: 
\small
\begin{subequations}
\label{zupdate2}
    \begin{flalign}
        z_{k+1}^v = &\argmin_{z^v} \biggl[ f^v(z^v)
        + \frac{\rho_R}{2} \| z^v - z^v_k + \bar{z}_k - \bar{\eta}_k + u_{k} \|_2^2 \biggr], && 
    \end{flalign}
    \begin{flalign}
         \qquad &\mathrm{s.t.} \qquad  \text{Separable Constraints } v, && \forall v\in\V,
    \end{flalign}
\end{subequations}
\normalsize
where $z^v$ is updated independently and in parallel by each vehicle.
\small
\begin{subequations}
\label{etaupdate2}
    \begin{flalign}
        \bar{\eta}_{k+1} = &\argmin_{\bar{\eta}} \biggl[g\left(|\V|\bar{\eta}\right) + \frac{|\V|\rho_R}{2}  \| \bar{\eta} - \bar{z}_{k+1} - u_{k} \|_2^2 \biggr], && 
    \end{flalign}
    \begin{flalign}
         \qquad &\mathrm{s.t.} \qquad  h\left(|\V|\bar{\eta}\right) \leq \omega, && 
    \end{flalign}
\end{subequations}
\normalsize
where $\bar{\eta}$ is the average of all local estimates $\eta^v$is update by the central controller. 
\small
\begin{flalign}
\label{uupdate2}
    u_{k+1} = u_{k} + \bar{z}_{k+1} - \bar{\eta}_{k+1}, &&
\end{flalign}
\normalsize
where $u$ is updated by the central controller. 

Finally, we clarify and further simplify the $\bar{\eta}$ update. Recall that $\eta^v$ is the local copy of the dispatch, so $\bar{\eta}$ is the vectorization of all \textit{shared} dispatch decisions ($Y$, $ep$, $eq$, $\hat{\gamma_{ch}}$, and $\hat{\gamma_{dch}}$). Since each of these appears independently in the constraints, the problem can be solved independently. As in the upper-level problem, $\rho_R$ can be split by variable type. We define $\rho^Y_R$, $\rho^N_R$, $\rho^p_R$, and $\rho^q_R$ as the penalty parameters associated with each decision variable type. Since vehicles share charging stations for both charging and discharging, both $\hat{\gamma}_{ch}$ and $\hat{\gamma}_{dch}$ must share $\rho^N_R$. Using this, we can split equation \eqref{etaupdate2} into three parts. 

First, constraint \eqref{N} states that the number of vehicles left to charge cannot exceed the number of charging stations. Then: 
\small
\begin{subequations}
\label{etaupdate3}
    \begin{flalign}
        \bar{\eta}^{\hat{\gamma}_{ch}}_{k+1}, \bar{\eta}^{\hat{\gamma}_{dch}}_{k+1} = \argmin_{\bar{\eta}^{\hat{\gamma}_{ch}}, \bar{\eta}^{\hat{\gamma}_{dch}}} &
         \left( \bar{\eta}^{\hat{\gamma}_{ch}} - \bar{z}_{k+1}^{\hat{\gamma}_{ch}} - u_{k}^{\hat{\gamma}_{ch}} \right)^2 \nonumber \\ &
        + \left( \bar{\eta}^{\hat{\gamma}_{dch}} - \bar{z}_{k+1}^{\hat{\gamma}_{dch}} - u_{k}^{\hat{\gamma}_{dch}} \right)^2, && 
    \end{flalign}
    \begin{flalign}
         \qquad &\mathrm{s.t.} \qquad  \bar{\eta}^{\hat{\gamma}_{ch}} + \bar{\eta}^{\hat{\gamma}_{dch}} \leq \frac{N}{|\V|}, && 
    \end{flalign}
\end{subequations}
\normalsize
where $\bar{\eta}^{\hat{\gamma}_{ch}}$ and $\bar{\eta}^{\hat{\gamma}_{dch}}$ must be updated jointly, but can be separated from the other variables and separated across each node $q\in \N_R$ and each timestep $\tau \in [0,\T]$. Note that $\rho^N_R$ must appear in equation \eqref{zupdate2} but can be dropped from \eqref{etaupdate3}. 

Following similar logic, each $\bar{\eta}^{ep}$ and $\bar{\eta}^{eq}$ can be solved for independently for each node $q\in \N_R$ and each timestep $\tau \in [0,\T]$:
\small
\begin{subequations}
\label{etaupdate4}
    \begin{flalign}
        \bar{\eta}^{ep}_{k+1} = &\argmin_{\bar{\eta}^{ep}} \biggl[g\left(|\V|\bar{\eta}^{ep}\right)+ \frac{|\V|\rho_R^p}{2}  \left( \bar{\eta}^{ep} - \bar{z}_{k+1}^{ep} - u_{k}^{ep} \right)^2 \biggr], && 
    \end{flalign}
\end{subequations}
\begin{subequations}
\label{etaupdate5}
    \begin{flalign}
        \bar{\eta}^{eq}_{k+1} = &\argmin_{\bar{\eta}^{eq}} \biggl[g\left(|\V|\bar{\eta}^{eq}\right)  + \frac{|\V|\rho_R^q}{2}  \left( \bar{\eta}^{eq} - \bar{z}_{k+1}^{eq} - u_{k}^{eq} \right)^2 \biggr]. && 
    \end{flalign}
\end{subequations}
\normalsize
Finally, we consider the passenger service portion of the dispatch, estimated in aggregate by $\bar{\eta}^{Y_{qr}}$. This is constrained to be less than the queue of waiting passengers but is also used to create the queue length estimates at future timesteps. We rewrite constraints \eqref{eqw} and \eqref{Equation_2} in one equation as: 
\small
\begin{align}
    \sum_{v \in \V} \sum_{\tau = 0}^{t^\prime}
    Y_{qr}^v(t_\tau) \leq& w_{qr}(t) + \sum_{\tau = 0}^{t^\prime} \tilde{d}_{qr}(t_\tau),  \nonumber \\ &\forall (q,r)\in\N_R^2, \forall t^\prime \in [0,\T],
	\label{omega}
\end{align}
\normalsize
where $t^\prime$ is an intermediate time between $t$ and $\T$. The right-hand side of equation \eqref{omega} is now a constant (since the initial queue length and exogenous demand estimates are known in advance). Then, $\bar{\eta}^{Y}$ can be updated as: 
\small
\begin{subequations}
\label{etaupdate6}
    \begin{flalign}
        \bar{\eta}^{Y_{qr}}_{k+1} = &\argmin_{\bar{\eta}^{Y_{qr}}} \biggl[\| \bar{\eta}^{Y_{qr}} - \bar{z}_{k+1}^{Y_{qr}} - u_{k}^{Y_{qr}} \|_2^2 \biggr], && 
    \end{flalign}
    \begin{flalign}
         \qquad &\mathrm{s.t.} \qquad  h\left(|\V|\bar{\eta}^{Y_{qr}}\right) \leq \omega_{qr}, && \label{omega2}
    \end{flalign}
\end{subequations}
\normalsize
where equation \eqref{omega2} comes directly from \eqref{omega}. Note that this is a joint decision for all timesteps but can be separated for each node $q\in \N_R$. As in the $\bar{\eta}^{\hat{\gamma}_{q,ch}}$ and $\bar{\eta}^{\hat{\gamma}_{q,dch}}$ update, $\rho^Y_R$ can be dropped from \eqref{etaupdate6}. 

ADMM can then be performed by first running the optimization program \eqref{zupdate2} independently and in parallel for each vehicle. The vehicles must report their dispatch decisions ($Y$, $ep$, $eq$, $\hat{\gamma_{ch}}$, and $\hat{\gamma_{dch}}$). That is, the central controller must know which passengers are being picked up, when and where vehicles are parked to charge, and the amount of charge being taken or returned to the grid. Any other information such as routing, state of charge, passenger payments, etc. can be kept private. Next, the central controller updates all $\bar{\eta}$ variables using equations \eqref{etaupdate3}, \eqref{etaupdate4}, \eqref{etaupdate5}, and \eqref{etaupdate6}. These can also be run in parallel, though they must be run by the central controller (not the individual vehicles). The results can the be sent back to the vehicles for the Lagrangian multiplier update \eqref{uupdate2} and the next iteration can proceed until convergence. 

\subsection{Convergence}
To reiterate, the ADMM procedures developed in this section have no guarantees of convergence for the MILPs they are intended to solve. However, appropriate choices of the ADMM penalty terms as well as some other minor modifications to the problem alow the proposed algorithm to achieve high-quality solutions.

We have already discussed that each penalty term $\rho$ can be separated based on variable type. This can be important when dealing with integer and continuous variables in the same problem. Since variables such as $Y^v_{rs}(t)$ can be heavily impacted by small fluctuations in objective costs (both spatially and temporally) due to the nature of the vehicle's trajectory, large values of $\rho_R^Y$ can cause very large changes in dispatch decisions across the entire fleet. On the other hand, given a stable dispatch solution, relatively values of $\rho_E^p$ are needed to get vehicles to discharge back to the grid. These concerns are very important when using ADMM as a heuristic for the MILP, so tuning is required to achieve fast and stable convergence to near-optimal solutions. 

The relationship between variables in a Markovian structure (such as the queuing model used in vehicle dispatch) can make the convergence of the ADMM algorithm harder. In general, for a single vehicle, an infeasibility at the end of the time horizon could require the entire route to be altered in the next iteration. However, in the model predictive control algorithm, solutions at future timesteps are discarded. Thus, it is less important to achieve either optimality or feasibility at future times (particularly if future energy and passenger demands are not well known). For this reason, it is helpful to modify the penalty parameters $\rho$ to be time-dependent. We define $\rho (t) = \frac{\rho}{(\tau + 1)^\alpha}$, where $\alpha$ is a small constant. This reduces the ADMM penalty over time, making large shifts to vehicle routing early in the time horizon less likely. 

The other issue that arises when using this approach is that many vehicles may appear to be identical. Though this concern may be alleviated if vehicles are heterogeneous, it is possible that SAEVs in the future could become homogeneous (as they are in the case study below). In that case, vehicles starting from the same node with similar states of charge will always behave in exactly the same way. This can lead to cycling behavior of the algorithm as vehicles jump between two infeasible points repeatedly. This behavior has been documented in other problems such as the unit commitment problem \citep{zhang_two-stage_2020}. Both cycling behavior and the issue of vehicle homogeneity can be solved by adding a proximal term $\frac{\epsilon}{2} \| z^v - z^v_k\|_2^2$. This is a special case of the term added by \citet{zhang_two-stage_2020}. However, instead of $(z^v - z^v_k)^TP(z^v - z^v_k)$, we use the $L_2$ norm. We also define $\epsilon$ as a vector, rather than a constant, where each element $\epsilon^v$ is drawn from a random distribution $U(0,\bar\epsilon)$. This serves to differentiate vehicles by their willingness to deviate from their current optimal solution. As the scaled dual variables change, the vehicle is only switches routes when it is substantially more beneficial, and these decisions are made at different times based on the value of $\epsilon^v$. Generally, this should lead to more stable convergence, while still achieving similar solutions. 

\section{Numerical Demonstrations}
In this paper, we examine the properties of our dispatch strategy on a small but realistic network (5-node distribution system and 5-node transportation system with 10 SAEVs found in \citet{robbennolt_resilience_2023}). This allows for comparison between the centralized solution and the decentralized ADMM solution. To solve the centralized problem (MILP) and the sub-problems of the decentralized problem (all MILPs or MIQPs) we use IBM ILOG CPLEX version 12.9. Additional tests are run on the larger Sioux Falls network (24 nodes) \citep{transportation_networks_for_research_core_team_transportation_nodate} using the IEEE-85 node network for the electric distribution system \citep{montoya_two-stage_2022} with 150 SAEVs. A simulation model was created in Java to test to long term behavior of vehicles and collect queuing and energy service information. Simulations were run on a laptop computer with an Intel Core i7-1165 at 2.80 GHz and 16 GB of RAM.

\begin{figure}[h]
	\centering
	\includegraphics[width=0.8\linewidth]{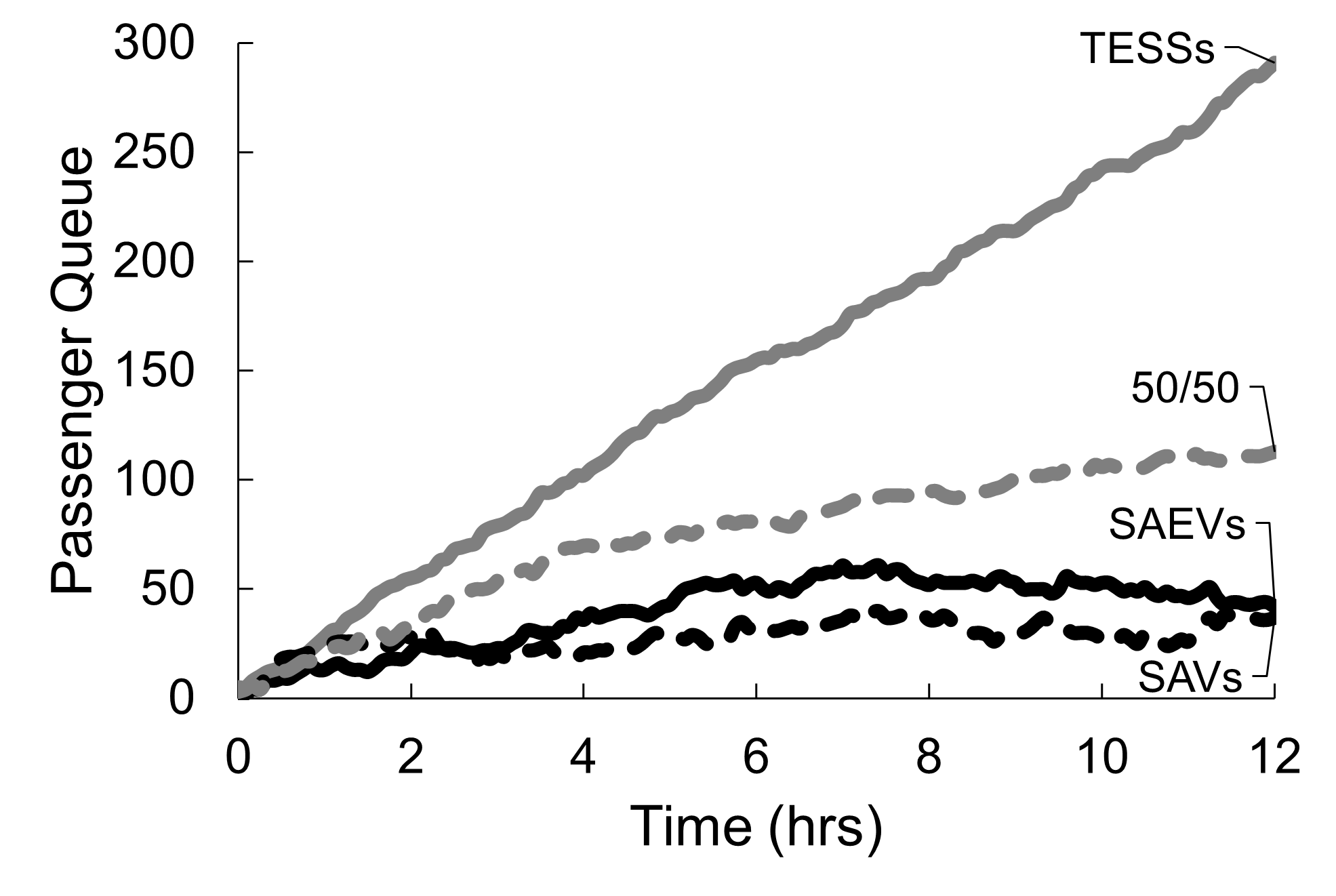}
	\caption{\small Cumulative unserved passengers for different vehicle fleets. Queues are only stable for fleets of SAEVs and SAVs.}
	\label{PassengerQueue}
\end{figure}

\begin{figure}[h]
	\centering
	\includegraphics[width=0.8\linewidth]{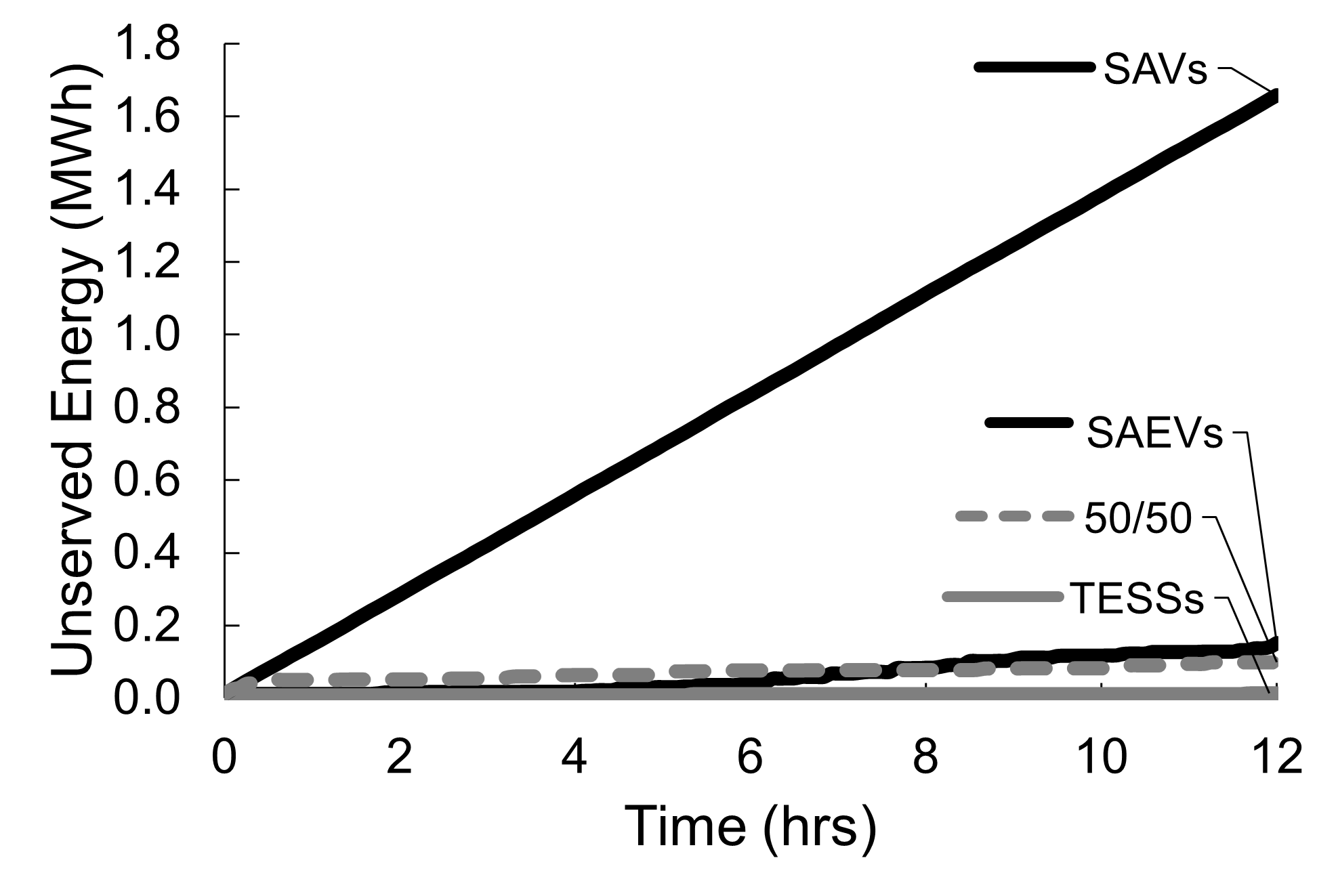}
	\caption{\small Cumulative unserved energy (MWh) for different vehicle fleets. SAEVs and 50/50 fleets are able to serve almost all energy demand that cannot be served by the grid.}
	\label{EnergyQueue}
\end{figure}

Passengers are loaded based on a Poisson distribution and electric demands are perturbed around the mean using a normal distribution (within 10\%). We consider a contingency scenario in which approximately 35\% (21\% for the IEEE 85-node network) of the energy demand cannot be served from the grid and must be moved by vehicles. We consider vehicles acting as SAVs that do not discharge to the grid (but must charge from the grid), TESSs that charge and discharge but do not serve passengers, and SAEVs dispatched using policy $\pi^\star$ (showcased in Figures \ref{PassengerQueue} and \ref{EnergyQueue}). For the toy network, we also consider a scenario in which both types of vehicles are present, half SAVs and half TESSs. For this case study all of the pricing functions $\mu$ are given a constant value ($\mu_0 = 1$ \$/pass, $\mu_1 = 20$ \$/hr, $\mu_2 = 500$ \$/MWh, $\mu_3 = 100$ \$/MWh, and $\mu_4 = 50$ \$/MWh). 

The experiments on the toy network demonstrate that the combination of SAV services and grid restoration does not significantly reduce passenger service performance. Figure \ref{PassengerQueue} shows that passenger queue lengths approximately double when vehicles also serve the grid. However, the service rate is less than the incoming demand rate (meaning in the long run all passengers will be served). In contrast, if only 5 vehicles serve passengers, the passenger queues will be unstable (eventually growing to infinity). 

When comparing the cumulative unserved energy, the TESSs are able to serve almost all the energy demand. When only half the vehicles are available or when they are also being used for passenger service the performance is comparable. In these cases, there are some timesteps when all demand cannot be served (usually due to stochasticity and unexpected increases in the demand). In both cases if the vehicles focus on one system exclusively there are detriments to the other. SAVs serving passengers alone neglect 35\% of the electric demand, and TESS serving passengers alone lead to unbounded passenger queues (and waiting times).

\begin{figure}[h]
	\centering
	\includegraphics[width=0.8\linewidth]{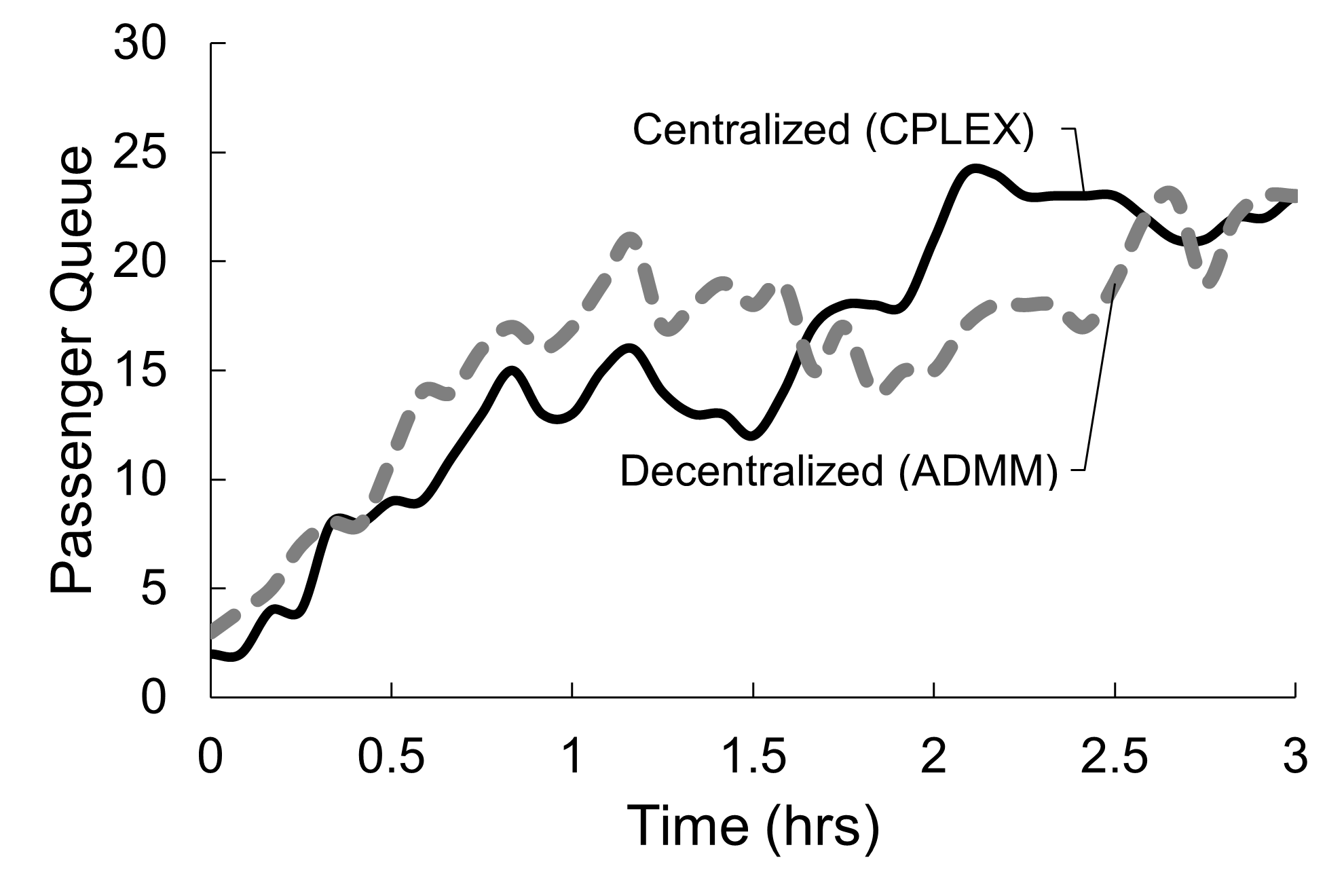}
	\caption{\small Validation of ADMM algorithm based on cumulative unserved passengers. The decentralized method achieves queues that are comparable to the centralized approach.}
	\label{QueueADMM}
\end{figure}

\begin{figure}[h]
	\centering
	\includegraphics[width=0.8\linewidth]{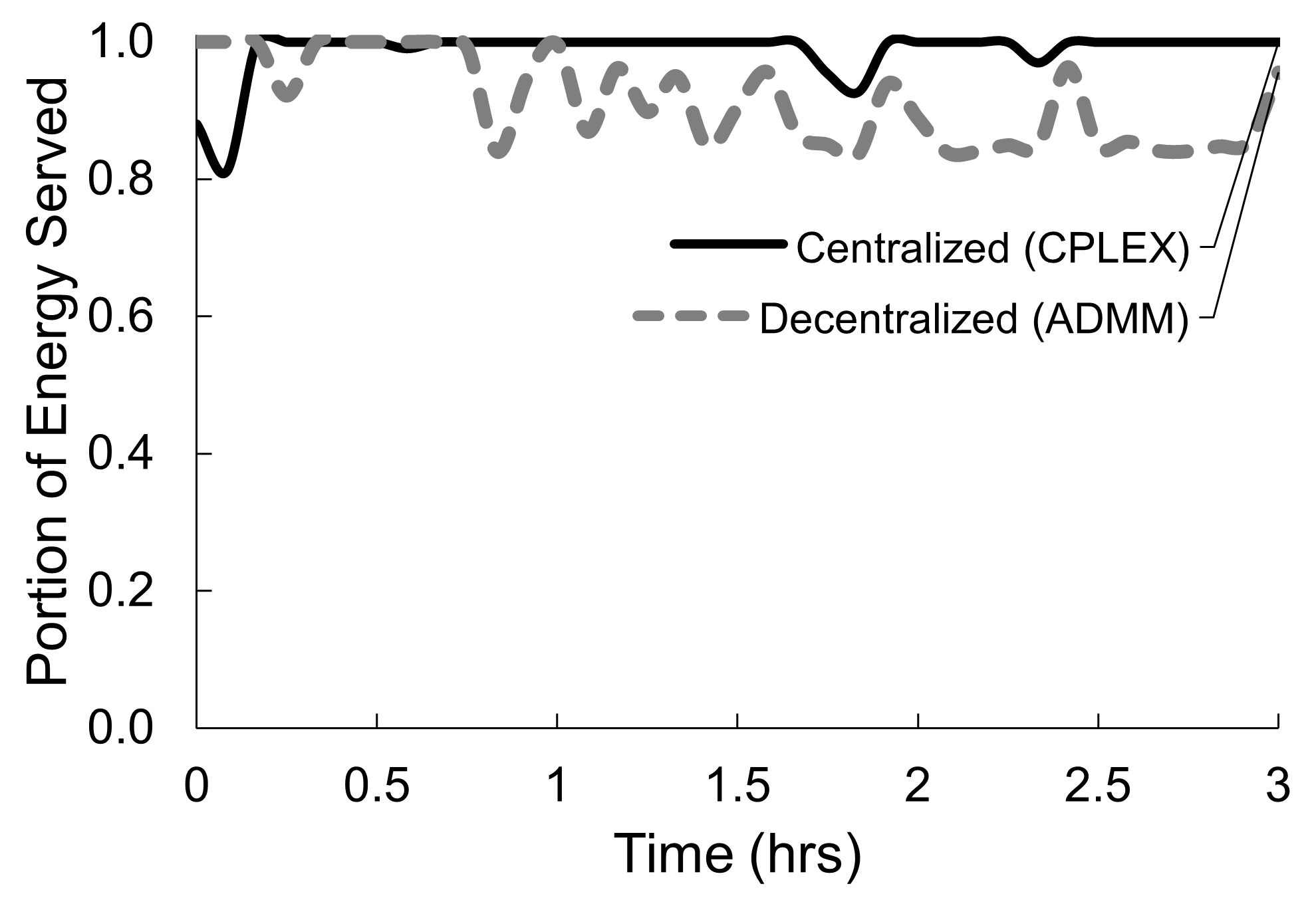}
	\caption{\small Validation of ADMM algorithm based on cumulative unserved energy (MWh). The decentralized policy sometimes serves slightly less energy but is still able to serve almost all the demand.}
	\label{EnergyADMM}
\end{figure}

We next compare our decentralized ADMM algorithm with the centralized version using CPLEX. Examining the SAEV dispatch policy over the first three hours of operation in Figures \ref{QueueADMM} and \ref{EnergyADMM}, we can observe the process of queue formation and energy service. After a simulation window of three hours (solving the dispatch problem every 5 minutes), both policies have stable passenger queues and serve approximately the same portion of the energy demand over time. These plots show that the decentralized policy is able to approximate near-optimal solutions without high levels of communication. Though the solutions each timestep are not always exactly the same (since the ADMM approach is a heuristic), the maximum stability dispatch ensures that these errors are more likely to be corrected at later timesteps \citep{kang_maximum-stability_2021}. Additional heuristics built into the end of the process, ensuring vehicles are discharging at higher rates to serve all demand could be explored in the future to speed the convergence of the algorithm and further reduce the difference between the two curves in Figure \ref{EnergyADMM}. 


\begin{figure}[h]
	\centering
	\includegraphics[width=\linewidth]{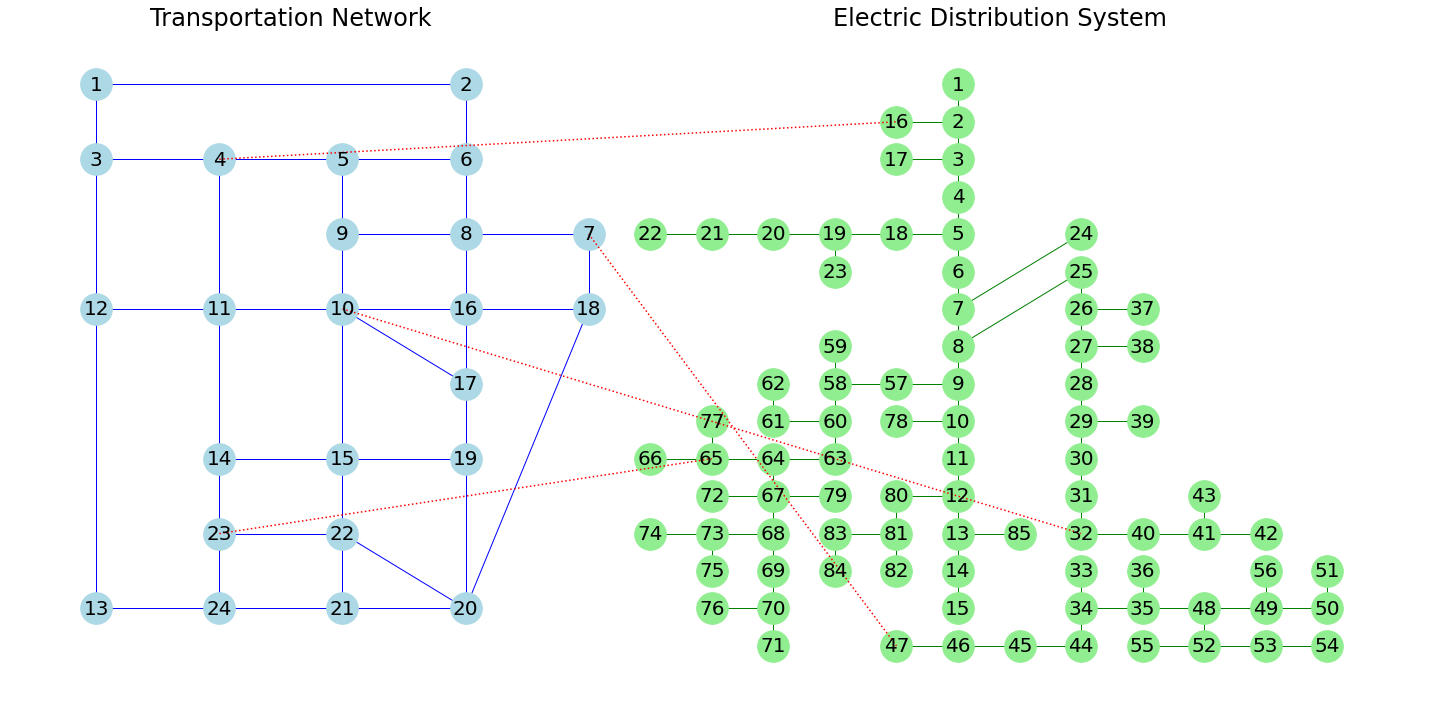}
	\caption{\small Sioux Falls and IEEE 85-node networks connected by charging stations (red).}
	\label{netfig}
\end{figure}

\begin{figure}[h]
	\centering
	\includegraphics[width=0.8\linewidth]{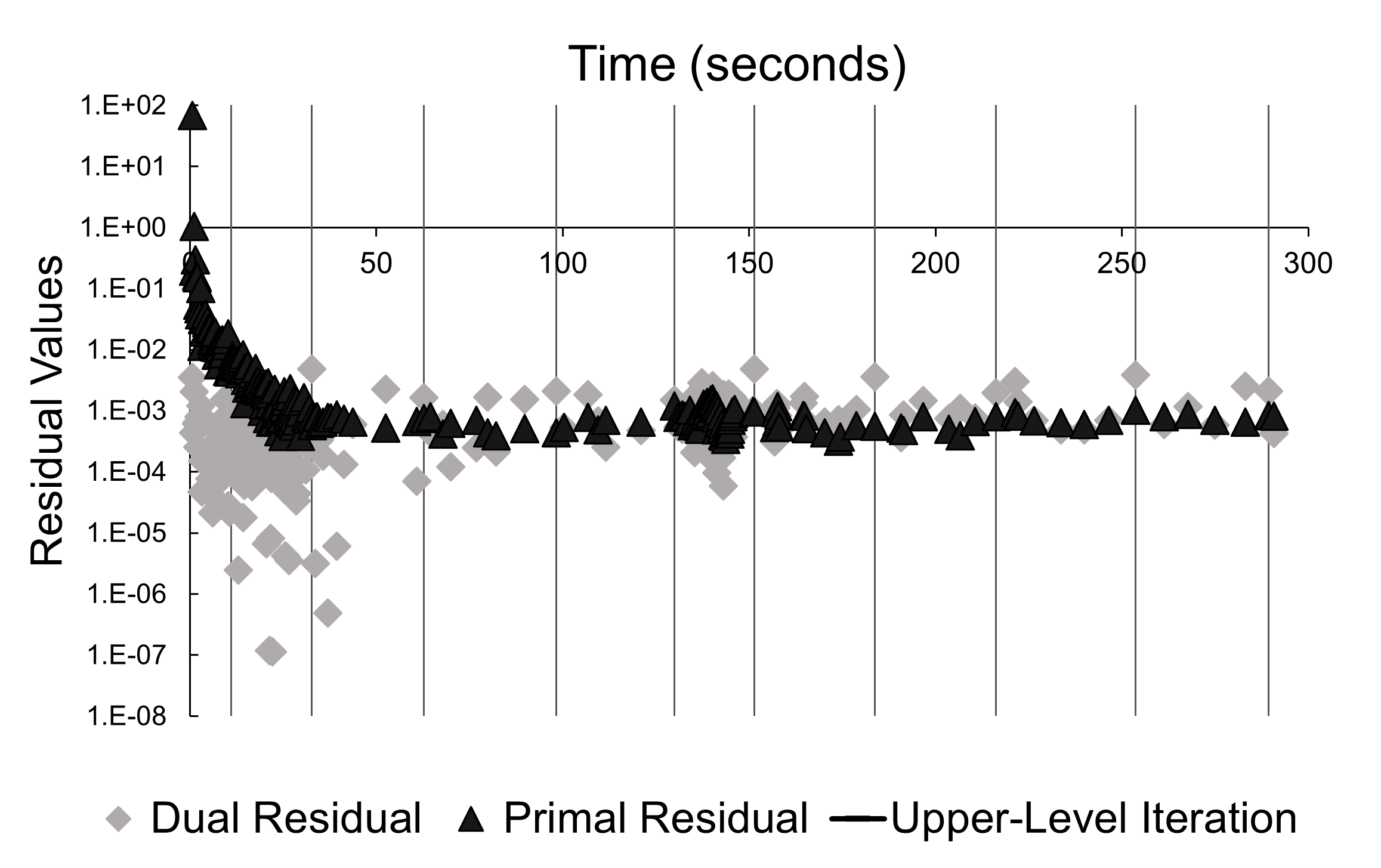}
	\caption{\small Validation of ADMM algorithm for the Sioux Falls network (points are lower-level iterations and lines are upper-level iterations.}
	\label{ADMM_Convergence_Fig}
\end{figure}

We run similar tests on the larger Sioux Falls transportation network with the IEEE 85-node distribution network. These networks are shown in Figure \ref{netfig} and are connected by four sets of charging stations. We test the operations of vehicles when used as SAV, TESSs, and SAEVs and there are two breaks in the electric grid between nodes 31--32 and nodes 34--44. Centralized solutions times range from about 10 minutes to several hours using CPLEX (if they can be found at all). Near-optimal solutions can be found in 1--5 minutes using the decentralized ADMM approach (assuming vehicles can optimize their own routes in parallel). Figure \ref{ADMM_Convergence_Fig} shows the convergence process of the algorithm on the larger Sioux Falls network for the first timestep. This is the most difficult timestep to solve since the few vehicles in the network mean that the greedy solution of each vehicle will lead to many conflicts. However, the primal and dual residuals both drop below $10^{-3}$ in about 60 seconds (about 4 upper-level iterations and 143 lower level iterations). The residuals could be reduced in the future or convergence rates could be increased with additional parameter tuning, though this may be unnecessary given the already fast convergence and the stable, near-optimal solution quality. Additional speed gains can be found in the future by developing heuristics for the lower-level routing problems.

\begin{figure}[h]
	\centering
	\includegraphics[width=0.8\linewidth]{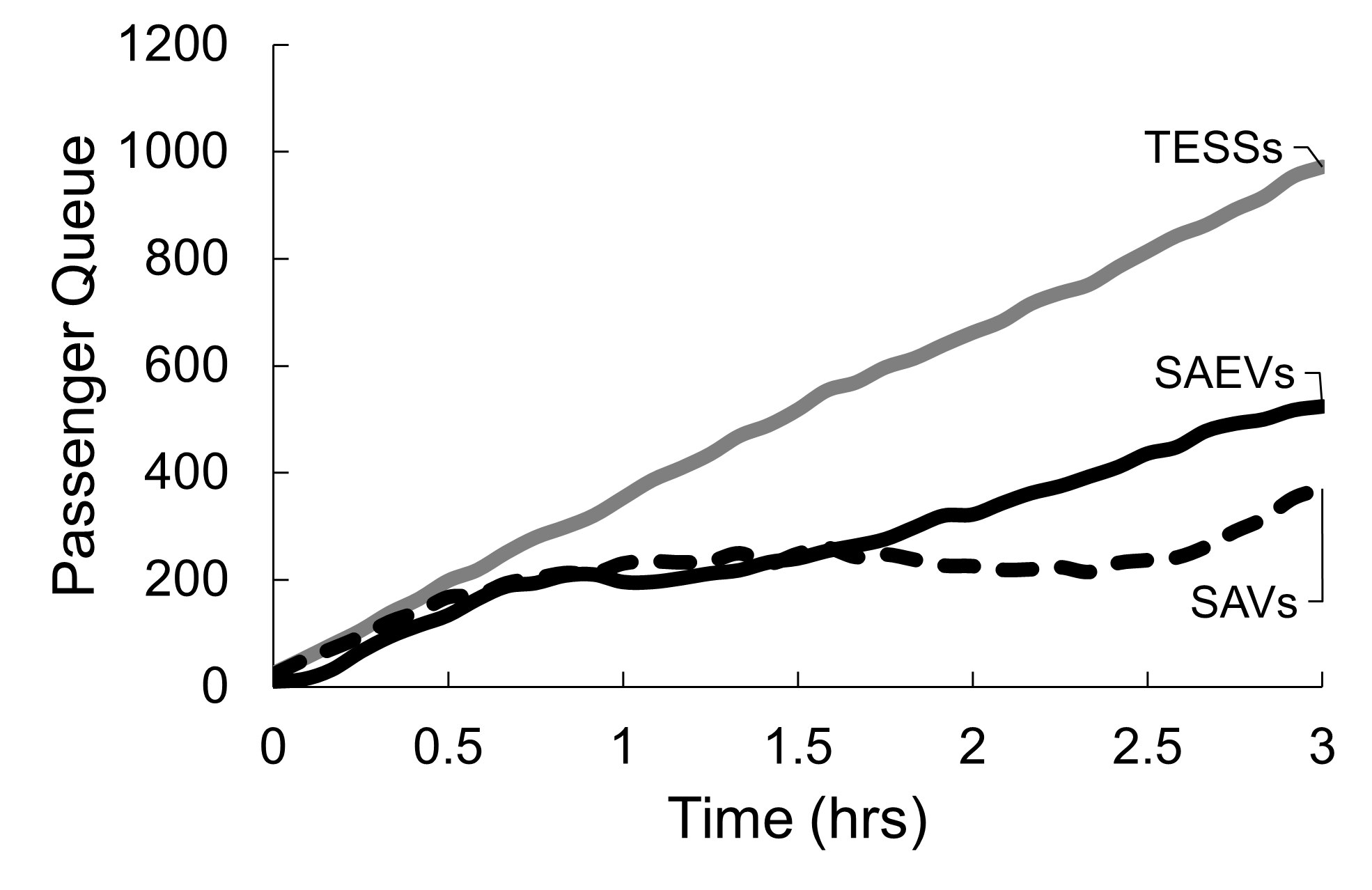}
	\caption{\small Cumulative unserved passengers for different vehicle fleets. SAEVs have slightly longer queues than SAVs, though demand starts to stabilize after 3 hours for both fleets.}
	\label{PassengerQueueLarge}
\end{figure}

\begin{figure}[h]
	\centering
	\includegraphics[width=0.8\linewidth]{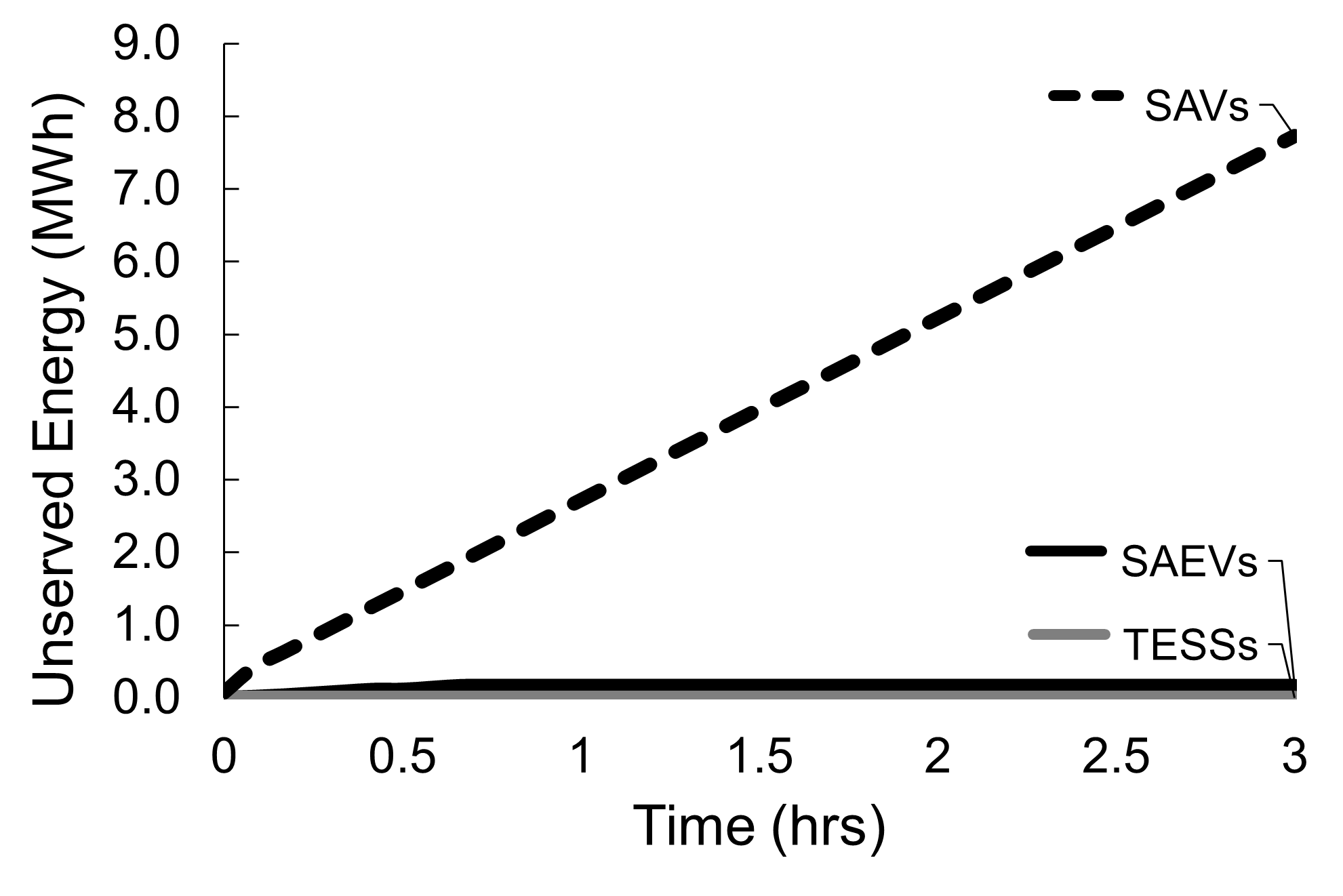}
	\caption{\small Cumulative unserved energy (MWh) for different vehicle fleets. Both SAEVs and TESSs serve almost all of the energy demand.}
	\label{EnergyQueueLarge}
\end{figure}

Figures \ref{PassengerQueueLarge} and \ref{EnergyQueueLarge} show a very similar pattern as for the smaller test network. The SAEVs are able to stabilize the demand with only slightly longer queues than if they were operated as SAVs, while also providing power to impacted areas. This is because vehicles already traveling between nodes carrying passengers can stop for a short time to discharge before carrying passengers the other direction and recharging. While neither fleet stabilizes passenger demands in the first 3 hours, the SAEV dispatch shows only slightly longer queues than that SAV fleet. 
These case studies suggest several implications for practical applications of SAEV routing methodologies and for future algorithmic improvements. This study suggests that ignoring interactions between SAEVs and the grid can be detrimental as charging and discharging behaviors add additional demands on the time of these vehicles. Ignoring such interactions could lead to underestimates of queue lengths as shown in Figures \ref{PassengerQueue} and \ref{PassengerQueueLarge}. On the other hand, ignoring the need to serve passengers (particularly during disaster scenarios) could lead to overestimates of the amount of power SAEVs can carry across outages (see figures \ref{EnergyQueue} and \ref{EnergyQueueLarge}). Though this paper focused on disaster scenarios, vehicles may also use vehicle to grid charging to reduce demand fluctuations, provide peak shaving, or many other grid services. However, demand fluctuations of energy and vehicles are linked, so this behavior needs to be better analyzed in a comprehensive framework to understand these impacts. 

When it comes to optimal dispatch, this work is built upon research into maximum throughput dispatch which ensures queues do not grow too large. The ADMM approach presented here actually presents some insight into this problem. When queues are very long, there are rarely conflicts when each vehicle takes its optimal path through the network (thus, computation times are relatively short). This aggregate behavior is exploited by previous approaches which do not need a time horizon and instead send vehicles to passengers with the longest head-of-line waiting time or with very short service times \citep{li_real-time_2021, xu_zone-based_2021, robbennolt_maximum_2023}. Though the SAEV dispatch proposed here relies on the time horizon, additional heuristics could be used to first route vehicles sequentially to achieve faster initial convergence to a feasible solution. When queues are shorter, there are more conflicts as many vehicles attempt to serve the same passengers (leading to much higher computation times). However, simple routing heuristics could be adopted in these cases since uncongested conditions are generally easier to solve with simple rules. Once the network becomes congested, the ADMM approach becomes increasingly easy to solve to near-optimal solutions as it also becomes increasingly valuable to find optimal routes. 

\section{Conclusions}
This paper examined the dispatch decision of SAEVs using the LinDistFlow model and a passenger queuing model. In the aftermath of a disaster these vehicles may be required to help serve critical electric loads while also providing mobility services for critical workers. While serving electric loads can improve resilience of the electric system, it can reduce the capacity of the vehicle fleet to serve passengers which could be detrimental for vulnerable populations. The model predictive control dispatch policy developed in this paper provides a framework for examining these competing objectives. 

We also provide a heuristic based on ADMM which provides fast, near-optimal solutions to the problem. This approach becomes increasingly effective as the network becomes congested (an important benefit as this is the time when optimal dispatch is most important). Using this approach, we developed two case studies which demonstrate that ignoring potential demands on these vehicles could lead to an overestimation of the benefits to one system or the other. On the other hand, well optimized dispatch still leads to substantial benefits to both systems. 

Future work should develop heuristics or use learning-based optimizers to solve the problem more quickly (for real-time implementation on realistic networks) and compare the ADMM heuristic to other solution approaches such as Benders decomposition. The lower-level problems of the ADMM heuristic were solved with CPLEX, though many exhibit structures that could be exploited in additional exact or heuristic solution approaches. Additional research should also examine the sensitivity to the cost functions, fleet size, charging/discharging rate, battery size, length of the time horizon, and behavior under different loading scenarios (i.e. more realistic demand profiles). Additional modifications to the optimization problem could include congestion created by the SAEV fleet or investigate the impacts of other energy storage solutions. Similarly, additional objective functions should be tested for day-to-day operation that could minimize demand fluctuations, prioritize green energy sources, incorporate equity concerns, etc. In addition to better understanding the SAEV routing problem, we believe that our proposed ADMM heuristic could be applicable to other vehicle routing approaches. This subproblem was shown to be very effective when the network was congested, and variants could be applied to other large-scale dispatch and routing problems. Finally, we have assumed a cooperative framework between the SAEV dispatcher and the power grid operator, possibly motivated by economic incentives or regulatory frameworks for post-disaster recovery. Future research should explore non-cooperative scenarios, potentially using game-theoretic frameworks to model strategic interactions and design appropriate incentive mechanisms.

\small
\bibliography{references}
\bibliographystyle{IEEEtranN}

\end{document}